\pdfoutput=1

\documentclass{article}

\usepackage[utf8]{inputenc}     
\usepackage[margin=10pt,font=small,labelfont=bf,labelsep=endash]{caption}
\usepackage[T1]{fontenc}
\usepackage{setspace}
\usepackage{booktabs}
\usepackage{multirow}
\usepackage{tabularx}
\usepackage{paralist}
\usepackage{ifthen}
\usepackage{listings}
\usepackage{amsmath}
\usepackage{amsfonts}
\usepackage{amssymb}
\usepackage{color}
\usepackage[table]{xcolor}
\usepackage{algorithm}
\usepackage{algpseudocode}
\usepackage{url}
\usepackage{rotating}
\usepackage{subfig}
\usepackage{arydshln}
\usepackage{todonotes}
\usepackage[normalem]{ulem}
\usepackage{siunitx}
\usepackage{diagbox}
\usepackage[htt]{hyphenat}
\usepackage{enumitem}
\usepackage{tikz}
\usepackage{pgfplots}
\usetikzlibrary{plotmarks}
\usepackage{authblk}
\usepackage{float} 
\usepackage{pdfpages}
\usepackage{csquotes} 
\usepackage[top=3cm, bottom=3cm, inner=3.5cm, outer=2.5cm]{geometry}
\usepackage[absolute]{textpos}

\newcommand{\swpackage}{\texttt}

\floatstyle{ruled}
\newfloat{proc}{thp}{lop}
\floatname{proc}{Procedure}

\makeatletter
\@ifpackageloaded{algorithm}
	{\algrenewcommand\algorithmicforall{\textbf{for each}}}
	{}
\@ifpackageloaded{amsmath}{
	\DeclareMathOperator*{\argmax}{arg\,max}
	\DeclareMathOperator*{\argmin}{arg\,min}}
	{}
\makeatother

\newcommand*\mean[1]{\overline{#1}}

\title{Model-independent comparison of simulation output}

\author[1]{Nuno Fachada}
\author[2]{Vitor V. Lopes}
\author[3]{Rui C. Martins}
\author[1]{Agostinho C. Rosa}

\affil[1]{Institute for Systems and Robotics, LARSyS, Instituto Superior Técnico, Universidade de Lisboa, Lisboa, Portugal}
\affil[2]{UTEC - Universidad de Ingeniería \& Tecnología, Lima, Jr. Medrano Silva 165, Barranco, Lima, Perú}
\affil[3]{INESC TEC, Campus da FEUP, Rua Dr. Roberto Frias, 4200-465 Porto, Portugal}

\providecommand{\keywords}[1]{\textbf{\textit{Keywords---}} #1}

\date{}

\begin{document}

\begin{textblock*}{210mm}(3mm,3mm)
\noindent The peer-reviewed version of this paper is published in Simulation Modelling Practice and Theory (\url{http://dx.doi.org/10.1016/j.simpat.2016.12.013}). This version is typeset by the authors and differs only in pagination and typographical detail.
\end{textblock*}

\maketitle

\begin{abstract}

Computational models of complex systems are usually elaborate and sensitive to implementation details, characteristics which often affect their verification and validation. Model replication is a possible solution to this issue. It avoids biases associated with the language or toolkit used to develop the original model, not only promoting its verification and validation, but also fostering the credibility of the underlying conceptual model. However, different model implementations must be compared to assess their equivalence. The problem is, given two or more implementations of a stochastic model, how to prove that they display similar behavior? In this paper, we present a model comparison technique, which uses principal component analysis to convert simulation output into a set of linearly uncorrelated statistical measures, analyzable in a consistent, model-independent fashion. It is appropriate for ascertaining distributional equivalence of a model replication with its original implementation. Besides model-independence, this technique has three other desirable properties: a) it automatically selects output features that best explain implementation differences; b) it does not depend on the distributional properties of simulation output; and, c) it simplifies the modelers' work, as it can be used directly on simulation outputs. The proposed technique is shown to produce similar results to the manual or empirical selection of output features when applied to a well-studied reference model.

\end{abstract}


\section{Introduction}

Complex systems are usually described as consisting of mutually interacting objects, often exhibiting complex global behavior resulting from the interactions between these objects. This behavior is typically characterized as ``emergent'' or ``self-organizing'' as the system's constituting parts do not usually obey a central controller \cite{boccara2004modeling}. Analytic treatment generally does not yield the complete theory of a complex system. Therefore, modeling and simulation techniques play a major role in our understanding of how these systems work \cite{bar1997dynamics}. Methodologies such as agent-based modeling (ABM), system dynamics, discrete event simulation, among others, are frequently employed for this purpose \cite{bagdasaryan2011discrete}. Of these, ABM provides an instinctive approach for describing many complex systems, as agents are regularly a suitable match to the individual and heterogeneous objects composing these systems. The local interactions of these objects, as well as their individual and adaptive behavior, are often critical for understanding global system response \cite{grimm2010odd,shirazi2014adaptive}. ABMs are commonly implemented as a stochastic process, and thus require multiple runs (observations) with distinct pseudo-random number generator (PRNG) seeds in order to have appropriate sample sizes for testing hypotheses and differentiating multiple scenarios under distinct parameterizations \cite{lee2015complexities}.

Computational models of complex systems in general, and ABMs in particular, are usually very sensitive to implementation details, and the influence that seemingly negligible aspects such as data structures, discrete time representation and sequences of events can have on simulation results is notable \cite{merlone2008}. Furthermore, most model implementations are considerably elaborate, making them prone to programming errors \cite{will2008replication}. This can seriously affect model validation\footnote{Determining if the model implementation adequately represents the system being modeled \cite{wilensky2007making} for its intended purpose \cite{robinson2004simulation}.} when data from the system being modeled cannot be obtained easily, cheaply or at all. Model verification\footnote{Determining if the model implementation corresponds to a specific conceptual model \cite{wilensky2007making}.} can also be compromised, to the point that wrong conclusions may be drawn from simulation results.

A possible answer to this problem is the independent replication of such models \cite{will2008replication}. Replication consists in the reimplementation of an existing model and the replication of its results \cite{thiele2015replicating}. Replicating a model in a new context will sidestep the biases associated with the language or toolkit used to develop the original model, bringing to light dissimilarities between the conceptual and implemented models, as well as inconsistencies in the conceptual model specification \cite{edmonds2003replication,wilensky2007making}. Additionally, replication promotes model verification, model validation \cite{wilensky2007making}, and model credibility \cite{thiele2015replicating}. More specifically, model verification is promoted because if two or more distinct implementations of a conceptual model yield statistically equivalent results, it is more likely that the implemented models correctly describe the conceptual model \cite{wilensky2007making}. Thus, it is reasonable to assume that a computational model is untrustworthy until it has been successfully replicated \cite{edmonds2003replication,david2013validating}.

Model parallelization is a an illustrative example of the importance of replication. Parallelization is often required for simulating large models in practical time frames, as in the case of ABMs reflecting systems with large number of individual entities \cite{fachada2015parallelization}. By definition, model parallelization implies a number of changes, or even full reimplementation, of the original model. Extra care should be taken in order to make sure a parallelized model faithfully reproduces the behavior of the original serial model. There are inclusively reports of failure in converting a serial model into a parallel one \cite{parry2012large}.

Although replication is considered the scientific gold standard against which scientific claims are evaluated \cite{peng2011reproducible}, most conceptual models have only been implemented by the original developer, and thus, have never been replicated \cite{axtell1996aligning,wilensky2007making,will2008replication,thiele2015replicating}. Several reasons for this problem have been identified, namely: a) lack of incentive \cite{will2008replication,thiele2015replicating}; b) below par communication of original models  \cite{peng2011reproducible,muller2014standardised}; c) insufficient knowledge and uncertainty of how to validate results of a reimplemented model \cite{wilensky2007making}; and, d) the inherent difficulty in reimplementing a model \cite{edmonds2003replication,wilensky2007making,will2008replication}. This work targets c), with positive influence on d). Replication is evaluated by comparing the output of the reimplementation against the output of the original model \cite{thiele2015replicating}, and this process, as will be discussed, is empirically driven and model-dependent (or even parameter-dependent). Furthermore, it is sometimes unclear as to what output features best describe model behavior. A robust and ready to use output comparison method would thus reduce or eliminate uncertainty of how to validate reimplementation results (reason c), eliminating this obstacle in the overall process of model replication (reason d).

We present a model comparison technique, which uses principal component analysis (PCA) \cite{jolliffe2002principal} to convert simulation output into a set of linearly uncorrelated statistical measures, analyzable in a consistent, model-independent fashion. It is appropriate for ascertaining statistical equivalence of a model replication with its original implementation. Besides model-independence, this technique has three additional desirable features: a) it automatically selects output features that best explain implementation differences; b) it does not depend on the distributional properties of simulation output; and, c) it simplifies the modelers' work, as it can be used directly on model output, avoiding manual selection of specific points or summary statistics. The proposed method is presented within the broader context of comparing and estimating the statistical equivalence of two or more model replications using hypothesis tests. The technique is evaluated against the empirical selection of output summary statistics, using the PPHPC ABM \cite{fachada2015template} as a test case. This model is thoroughly studied in terms of simulation output for a number of parameter configurations, providing a solid base for this discussion.

The paper is organized as follows. First, in Section~\ref{sec:background}, we review commonly used methods for comparing the output of simulation models, as well as previous work on model replication using these methods. An overview of hypothesis testing within the scope of model output comparison is conducted in Section~\ref{sec:tests}. The main steps in designing and performing a model comparison experiment using hypothesis tests are presented in Section~\ref{sec:doe}. The proposed model-independent comparison methodology is described in Section~\ref{sec:micomp}. Section~\ref{sec:simmod} introduces PPHPC, the simulation model used as a test case for the proposed model comparison approach. Section~\ref{sec:expsetup} delineates the experimental setup for assessing this methodology against the manual selection of output summary measures. In Section~\ref{sec:results}, results of the empirical and model-independent comparison approaches are presented. A discussion and an evaluation of these results is performed in Section~\ref{sec:discussion}. Recommendations on using the proposed method are given in Section~\ref{sec:recomend}. Section~\ref{sec:conclusions} summarizes what was accomplished in this paper.

\section{Background}
\label{sec:background}

Axtell et al. \cite{axtell1996aligning} defined three replication standards (RSs) for the level of similarity between model outputs (Carley \cite{carley2002computational} calls the RS the emphasis of demonstration): \textit{numerical identity}, \textit{distributional equivalence} and \textit{relational alignment}. The first, \textit{numerical identity}, implies exact numerical output, but it is difficult to demonstrate for stochastic models and not critical for showing that two such models have the same dynamic behavior. To achieve this goal, \textit{distributional equivalence} is a more appropriate choice, as it aims to reveal the statistical similarity between two outputs. Finally, \textit{relational alignment} between two outputs exists if they show qualitatively similar dependencies with input data, which is frequently the only way to compare a model with another which is inaccessible (e.g., implementation has not been made available by the original author), or with a non-controllable ``real'' system (such as a model of the human immune system \cite{Fachada2009}). For the remainder of this text we assume the \textit{distributional equivalence} RS when discussing model replication.

For the \textit{distributional equivalence} RS, a set of statistical summaries representative of each output are selected. It is these statistical summaries, and not the complete outputs, that will be compared in order to assert the similarity between the original model and the replication. As models may produce large amounts of data, the summary measures should be chosen as to be relevant to the actual modeling objective. The summaries of all model outputs constitute the set of focal measures (FMs) of a model \cite{wilensky2007making}, or more specifically, of a model parameterization (since different FMs may be selected for distinct parameterizations). There are three major statistical approaches used to compare FMs: 1) statistical hypothesis tests; 2) confidence intervals; and, 3) graphical methods \cite{balci1984validation}.

Statistical hypothesis tests are often used for comparing two or more model implementations \cite{axtell1996aligning,wilensky2007making,edmonds2003replication,miodownik2010between,radax2010prospects,fachada2015parallelization}. More specifically, hypothesis tests check if the statistical summaries obtained from the outputs of two (or more) model implementations are drawn from the same distribution. However, statistical testing may not be the best option for comparing the output of a model with the output of the system being modeled. Since the former is only an approximation of the latter, the null hypothesis that FMs from both are drawn from the same distribution will almost always be rejected. Although statistical procedures for comparing model and system outputs using hypothesis tests 
have been proposed \cite{sargent2010new}, confidence intervals are usually preferred for such comparisons, as they provide an indication of the magnitude by which the statistic of interest differs from model to system. Confidence intervals are also commonly used when evaluating different models that might represent competing system designs or alternative operating policies \cite{balci1984validation,law2014simulation}. Graphical methods, such as Q\textendash{}Q plots, can also be employed for comparing output data, though their interpretation is more subjective than the previous methods.

A number of simulation models have been replicated and/or aligned by using statistical methods to compare FMs. 
In reference \cite{axtell1996aligning}, Axtell et al. compared two initially different models, with one iteratively modified in order to be aligned with the other. The authors evaluated how can distinct equivalence standards be statistically assessed using non-parametric statistical tests (namely the Kolmogorov\textendash{}Smirnov \cite{massey1951kolmogorov} and Mann\textendash{}Whitney \textit{U} \cite{gibbons2011nonparametric} tests), and how minor variations in model design affect simulation outcomes. They concluded that comparing models developed by different researchers and with different tools (i.e., programming languages and/or modeling environments), can lead to exposing bugs, misinterpretations in model specification, and implicit assumptions in toolkit implementations. The concepts and methods of ``computational model alignment'' (or ``docking'') were first discussed in this work.

Edmonds and Hales \cite{edmonds2003replication} performed two independent replications of a previously published model involving co-operation between self-interested agents. Several shortcomings were found in the original model, leading the authors to conclude that unreplicated simulation models and their results cannot be trusted. This work is one of the main references in model replication, describing in detail the process of running two model implementations with different parameters, selecting comparison measures and performing adequate statistical tests.

In reference \cite{wilensky2007making}, the authors presented an ABM replication case study, describing the difficulties that emerged from performing the replication and determining if the replication was successful. A standard $t$-test was used for comparing model outputs. The authors concluded that model replication influences model verification and validation and promotes shared comprehension concerning modeling decisions.

Miodownik et al. \cite{miodownik2010between} replicated an ABM of social interaction \cite{bhavnani2003adaptive}, originally implemented in MATLAB, using the PS-I environment for ABM simulations \cite{dergachev2013psi}. A statistical comparison of the mean levels of ``civicness'' at the end of the simulation (over 10 runs) was performed using the Mann\textendash{}Whitney \textit{U} test. Results showed that, while distributional equivalence was obtained in some cases, the two models were mostly only relationally aligned. The authors attribute this mainly to the fact that some aspects of the original model were not implementable with PS-I.

A method for replicating insufficiently described ABMs was discussed in reference \cite{radax2010prospects}, consisting in modeling ambiguous assumptions as binary parameters and systematically applying statistical tests to all combinations for their equivalence to the original model. The approach was used to replicate Epstein's demographic prisoner's dilemma model \cite{epstein2008prisoners}, with only partial success, suggesting the existence of some undefined assumptions concerning the original model. The authors also conducted a number of statistical tests regarding the influence of specific design choices, highlighting the importance that these be explicitly documented. 

Alberts et al. \cite{alberts2012data} implemented a CUDA \cite{cuda2014} version of Toy Infection Model \cite{an2007toy}, and compared it with the original version implemented in NetLogo \cite{wilensky1999}, as well as to an in-house optimized serial version. Statistical validation was performed visually using Q\textendash{}Q plots. 

Multiple serial and parallel variants of the PPHPC model were compared in reference \cite{fachada2015parallelization}. Simultaneous comparison of the several variants was accomplished by applying the multi-group non-parametric Kruskal\textendash{}Wallis test \cite{kruskal1952use} to predetermined FMs of the several model outputs. Results showed that all model variants could not be considered misaligned for a number of different parameters.

\section{Statistical hypothesis tests for comparing FMs}
\label{sec:tests}

As described in the previous section, hypothesis tests are commonly used for comparing FMs collected from two or more model implementations. Since statistical tests are a central subject matter in this discussion, the current section aims to clarify their use in the context of model comparison. More specifically, this section illustrates: a) how to generically interpret statistical tests; b) what tests to use and how their choice strongly depends on the statistical properties of FMs; c) what techniques are available to assess these properties; and, d) how to deal with the case of multiple FMs and/or test results.

In statistical hypothesis testing, a null hypothesis ($H_0$) is tested against an alternative hypothesis ($H_1$). The test can have two results: 1) fail to reject $H_0$; or, 2) reject $H_0$ in favor of $H_1$. In the most simple terms, the result of a statistical test is derived from the $p$-value it generates. The $p$-value is the probability of obtaining a result equal or more unexpected than what was observed in the data if $H_0$ is true. If the $p$-value is below a predefined significance level $\alpha$, the test result is deemed statistically significant and the null hypothesis, $H_0$, is rejected.  However, the test result may not be clear, especially when the $p$-value is close to the typically used significance levels, $\alpha=0.01$ or $\alpha=0.05$. Thus, it may be preferable to show the $p$-value instead of presenting the test result as a binary decision.

The rejection of a true $H_0$ is designated as a false positive or a type I error. Consequently, it stems from the $p$-value definition that the significance level, $\alpha$, also represents the type I error rate. A type II error, or false negative, occurs when the test fails to reject a false $H_0$. The type II error rate, denoted as $\beta$, has an inverse relation with $\alpha$, such that there is a trade-off between the two. The statistical power of a test, defined as $1-\beta$, is the probability that it correctly rejects a false $H_0$. Reducing the type I error rate, $\alpha$, increases $\beta$, and consequently reduces the power of a test. Increasing the sample size is a general way to increase the power of a test without modifying $\alpha$ \cite{montgomery2010applied}.

In the context of model comparison, the tests of interest are two-sample (or multi-sample) statistical tests which test for the null hypothesis that the observations in each sample are drawn from the same distribution, against the alternative that they are not. Here, samples or groups correspond to the model implementations being compared. Typically, when more than one FM is to be compared, univariate tests are used to compare individual FMs. Nonetheless, FMs may also be combined into one multidimensional random variable, and then simultaneously compared using a multivariate test. In either case, the choice of test also depends on characteristics of the FM and on the assumptions that can be made concerning its underlying distribution. Tests which make such assumptions are labeled as parametric tests. Such tests are generally more powerful than their non-parametric counterparts, which do not expect any particular underlying distribution. Table~\ref{tab:stattests} lists tests which, according to these aspects, are commonly used to check if samples are drawn from the same distribution.

\begin{table}[ht] 
	\centering
	\begin{tabular}{p{2.6cm}p{3.2cm}p{3.9cm}}
		
		& Parametric & Non-parametric \\
		\toprule
		\parbox[t]{2.6cm}{Univariate\\($s=2$ groups)} & $t$-test \cite{montgomery2010applied} & \parbox[t]{3.9cm}{Mann\textendash{}Whitney \cite{gibbons2011nonparametric} \\ Kolmogorov\textendash{}Smirnov \cite{massey1951kolmogorov}} \\
		\midrule
		\parbox[t]{2.6cm}{Univariate\\($s>2$ groups)} & ANOVA \cite{montgomery2010applied} & Kruskal\textendash{}Wallis \cite{kruskal1952use} \\
		\midrule
		Multivariate & MANOVA \cite{krzanowski1998,tabachnick2013using} & Various \cite{anderson2001new,clarke1993nonparametric,mielke1976multi,rosenbaum2005exact,baringhaus2004new} \\
		\bottomrule
	\end{tabular}
	\caption{\label{tab:stattests}Hypothesis tests commonly used to check if samples are drawn from the same distribution. Tests are organized by parametric assumptions, sample dimensionality and number of groups (samples) which can be compared.}
\end{table}

The tests listed in Table~\ref{tab:stattests} expect that samples are mutually independent. This can be guaranteed if samples are properly collected, as discussed later in Section \ref{sec:doe:runs}. The parametric tests make more stringent assumptions, though, also requiring that \cite{boneau1960effects,balci1982validation,krzanowski1998}: a) each sample is drawn from a normally distributed population (multivariate normality for MANOVA); and, b) samples are drawn from populations with equal variances (for MANOVA, the variance\textendash{}covariance matrix should be the same for each population). These assumptions can be checked using additional statistical tests. More specifically, group sample normality can be assessed using the Shapiro\textendash{}Wilk test \cite{shapiro1965analysis} (Royston test \cite{royston1982extension} for multivariate samples), while equality of variances among groups can be verified with the Bartlett test \cite{bartlett1937properties} (Box's M test \cite{box1949general} for homogeneity of variance\textendash{}covariance matrices in the multivariate case). However, Box's M test is very sensitive and can lead to false negatives (type II errors). Fortunately, MANOVA is considerably robust to violations in the homogeneity of variance\textendash{}covariance matrices when groups have equal sample sizes \cite{tabachnick2013using}.

If these assumptions are not met, non-parametric tests may be used instead. Non-parametric tests may be also preferable if: a) a specific FM is better represented by the median (the listed parametric tests check for differences in means); b) the sample size is small; c) the data is not continuous; or, d) the data contains outliers.

If the choice falls on multivariate tests there are a few caveats. For example, in MANOVA each FM is a dependent variable (DV). However, MANOVA is not appropriate when: a) DVs are highly correlated, which may be the case for FMs derived from outputs of the same simulation run; and, b) when the number of DVs or dimensions is higher than the number of observations (i.e., when there are more FMs than runs per model implementation). Additionally, the non-parametric multivariate alternatives (e.g., \cite{anderson2001new,clarke1993nonparametric,mielke1976multi,rosenbaum2005exact,baringhaus2004new}) are not as widespread and well established as MANOVA, and are commonly oriented towards specific research topics.

When more than one FM is to be compared and the choice of test falls on univariate tests, several $p$-values will be generated, as one test will be applied per FM. This is known as a multiple comparisons problem. Since the $p$-value is the probability of obtaining a result equal or more unexpected than what was observed if $H_0$ is true, the generation of multiple $p$-values increases the likelihood of false positives, i.e., of rejecting a true $H_0$ in some of the performed tests. For example, with $\alpha=0.05$, it is expected that approximately 5\% of the tests incorrectly reject $H_0$. This problem can be addressed with a multiple testing correction procedure, such as the Bonferroni or Holm methods \cite{shaffer1995multiple}. However, multiple comparison correction methods often assume independency of test statistics, which might not be possible to assure when testing different outputs of the same simulation model, which are most likely correlated. Additionally, such approaches may increase the likelihood of type II errors. As such, it is often preferable to simply present the $p$-values and discuss possible interpretations, such that the reader can draw his own conclusions \cite{perneger1998s}.

In the case of models with multiple outputs and multiple statistical summaries per output, there may exist an intermediate logical grouping of FMs. More specifically, FMs can be logically grouped by output. In such cases, does the researcher ignore this logical grouping of FMs, and bundles them together in a multivariate or multiple univariate comparison problem? Or does the researcher treat FMs from different outputs separately? This a model-dependent issue, and generally, the simplest course of action is to perform a univariate test per FM, present all the (uncorrected) $p$-values, and discuss the results considering the increased likelihood of type I errors.

\section{Designing and performing a model comparison experiment}
\label{sec:doe}

In order to setup a model comparison experiment it is necessary to first define a research question, which can be stated as follows:

\begin{displayquote}
Determine if two or more implementations of a conceptual simulation model display statistically indistinguishable behavior.
\end{displayquote}

If two or more implementations produce the same dynamic behavior it is reasonable to expect that any output collected over a number of independent runs also comes from the same distribution \cite{edmonds2003replication}. Thus, the research question can be reformulated in the following manner:

\begin{displayquote}
Determine if two or more implementations of a conceptual simulation model generate statistically indistinguishable outputs, given the same inputs.
\end{displayquote}

As described in Section~\ref{sec:background}, output similarity is assessed using statistical summaries representative of each output, i.e., by verifying if the selected FMs are distributionally equivalent. Furthermore, to answer the original question with a higher degree of confidence, output similarity should be observed for distinct input parameter sets \cite{edmonds2003replication}. Consequently, the research question can be further refined:

\begin{displayquote}
Determine if two or more implementations of a conceptual simulation model generate statistically indistinguishable FMs for several parameterizations.
\end{displayquote}

Procedure \ref{proc:modcompexp} summarizes the process of performing a model comparison experiment using hypothesis tests to answer this research question. In the following subsections each step of the procedure is analyzed in detail.

\begin{proc}
\caption{Steps for performing a model comparison experiment.}
\label{proc:modcompexp}

\begin{enumerate}
	\item Choose the parameter sets with which to generate FMs.
	\item For each parameter set:
	\begin{enumerate}[label*=\arabic*]
		\item \label{proc:modcompexp:fepar:nrep} Perform $n$ replications for each model implementation and collect the respective outputs.
		\item \label{proc:modcompexp:fepar:selfms} Select FMs to compare.
		\item \label{proc:modcompexp:fepar:extfms} Extract FMs from collected outputs.
		\item \label{proc:modcompexp:fepar:hyptests} Analyze statistical properties of FMs.
		\item \label{proc:modcompexp:fepar:compfms} Compare FMs using hypothesis tests.
		\item \label{proc:modcompexp:fepar:decide} Decide on the alignment or otherwise of the compared model implementations for the current parameter set.
	\end{enumerate}
	\item Decide on the global alignment or otherwise of the compared model implementations.
\end{enumerate}

\end{proc}

\subsection{Choose the parameter sets with which to generate FMs}
\label{sec:doe:defineparamsets}

In order to demonstrate that two or more model implementations are misaligned, it is sufficient to show that at least one FMs is statistically different for one parameterization. On the other hand, demonstrating that FMs are drawn from the same distribution for any number of parameter sets does not definitively prove that the compared implementations are aligned. Nonetheless, demonstrating alignment for more parameter sets increases the confidence that the implementations are in fact globally aligned \cite{edmonds2003replication}.

As a starting point, a minimum of two parameterizations should be used to verify if implementations are aligned. The chosen parameter sets should induce noticeably distinct simulation behaviors, thus triggering different model mechanisms and increasing the probability of finding implementation differences.

\subsection{Perform a number of replications for each model implementation and collect the respective outputs}
\label{sec:doe:runs}

Statistical best practices should be followed when setting up the replications or runs for each model implementation. More specifically: a) there should be enough replicates in order to guarantee adequate statistical power; and, b) the replicates should be independently generated.

Considering the first issue, an appropriate minimum sample size (i.e., number of replicates or runs per model implementation) is required so that the test adequately rejects false null hypotheses.  Long run times in more complex models may limit the number of observations which can be collected in practice. While small sample sizes reduce the likelihood that statistically significant results indicate a true difference, large samples can lead to the detection of small and/or irrelevant differences \cite{montgomery2010applied}. The determination of sample size for a desired power level depends on the distributional properties of each FM, which in practice could mean that different FMs warrant distinct samples sizes. Additionally, in a two-sample or $s>2$ sample testing scenario, it is preferable to have equal numbers of observations from the populations being compared, since this may facilitate interpretation of results and make parametric tests more robust to variance homogeneity assumptions \cite{tabachnick2013using,montgomery2010applied}. As such, sample size should be chosen by balancing the desired statistical power, the available computational resources, as well as the simplicity and/or convenience of the analysis to be performed. A thorough discussion on determining the minimum sample size for a desired power level in the context of simulation models is presented in reference \cite{lee2015complexities}. 

The second issue concerns the independence of replicates. This can be achieved by performing each run with a distinct pseudo-random number stream. Typically, the same PRNG algorithm with different seeds is used for this purpose. However, some care is required in order to avoid correlated or overlapped streams. For example, using the run number as the seed should be avoided.  A simple solution consists of using a random spacing approach to split the PRNG stream, for example by applying a cryptographic hash (distinct from the PRNG algorithm used in the simulations) to the run number, and use the resulting value as the PRNG seed \cite{fachada2015template,fachada2015parallelization}. Allying this technique with a long-period PRNG such as the Mersenne Twister \cite{matsumoto1998mersenne} minimizes the likelihood of correlated or overlapped streams being used in different runs.

\subsection{Select FMs to compare}
\label{sec:doe:definefms}

Output summary measures, or more generally, FMs, are commonly selected in an empirical fashion, and normally consist of long-term or steady-state means. Nonetheless, being limited to analyze average system behavior can lead to incorrect conclusions \cite{law2014simulation,lee2015complexities}. Consequently, other measures such as proportions or extreme values can also be used to assess model behavior.

For steady-state FMs, care must be taken with initialization bias, which may cause substantial overestimation or underestimation of the long-term performance \cite{sanchez1999abc}. Such problems can be avoided by discarding data obtained during the initial transient period, before the system reaches steady-state conditions. The simplest way of achieving this is to use a fixed truncation point, $l$, for all runs with the same initial conditions, selected such that: a) it systematically occurs after the transient state; and, b) it is associated with a round and clear value, which is easier to communicate \cite{sanchez1999abc}. A moving average procedure for selecting $l$ is discussed in references \cite{fachada2015template,law2014simulation}.

\subsection{Extract FMs from collected outputs}
\label{sec:doe:extractfms}

Let $\mathbf{X}_j=\begin{bmatrix} X_{j0} & X_{j1} & X_{j2} & \ldots & X_{jm} \end{bmatrix}$ be an output collected from the $j^{\text{th}}$ simulation run of one model implementation. The $X_{ji}$'s (rows under `Iterations' in Table~\ref{tab:runstats}) are random variables that will, in general, be neither independent nor identically distributed \cite{law2014simulation}, and as such, are not adequate to be used directly in a classical statistical analysis. On the other hand, let $X_{1i}, X_{2i}, \ldots, X_{ni}$ be the observations of an output at iteration $i$ for $n$ runs (columns under `Iterations' in Table~\ref{tab:runstats}), where each run begins with the same initial conditions, but uses a different stream of random numbers as a source of stochasticity. The $X_{ji}$'s will now be independent and identically distributed (IID) random variables, to which classical statistical analysis can be applied. However, because individual values of the output at some iteration $i$ are unlikely to be representative of the output as a whole, summary measures, as shown in Table~\ref{tab:runstats}, under `Statistical summaries', should be used instead.

\begin{table}[ht]
	\centering
	\resizebox{\columnwidth}{!}{%
		\begin{tabular}{c|ccccc|ccccc}
			\toprule
			Rep. & \multicolumn{5}{c|}{Iterations} & \multicolumn{5}{c}{Statistical summaries} \\
			\midrule
			1    & $X_{10}$ & $X_{11}$ & $\ldots$ & $X_{1,m-1}$ & $X_{1,m}$ & $f_1(\mathbf{X}_{1})$ & $f_2(\mathbf{X}_{1})$ & $\ldots$ & $f_{q-1}(\mathbf{X}_{1})$ & $f_q(\mathbf{X}_{1})$ \\
			2    & $X_{20}$ & $X_{21}$ & $\ldots$ & $X_{2,m-1}$ & $X_{2,m}$ & $f_1(\mathbf{X}_{2})$ & $f_2(\mathbf{X}_{2})$ & $\ldots$ & $f_{q-1}(\mathbf{X}_{2})$ & $f_q(\mathbf{X}_{2})$ \\
			$\vdots$ & $\vdots$ & $\vdots$ & $\ddots$ & $\vdots$ & $\vdots$ & $\vdots$ & $\vdots$ & $\ddots$ & $\vdots$ & $\vdots$ \\
			$n$  & $X_{n0}$ & $X_{n1}$ & $\ldots$ & $X_{n,m-1}$ & $X_{n,m}$ & $f_1(\mathbf{X}_{n})$ & $f_2(\mathbf{X}_{n})$ & $\ldots$ & $f_{q-1}(\mathbf{X}_{n})$ & $f_q(\mathbf{X}_{n})$ \\
			\bottomrule
		\end{tabular}}
		\caption{\label{tab:runstats}Values of a generic simulation output from one model implementation (under `Iterations') for $n$ replications of $m$ iterations each (plus iteration 0, i.e., the initial state), and the respective summary measures (under `Statistical summaries'). Values along columns are IID.}
\end{table}
	
The statistical summaries, or more generally, FMs, can be empirically defined as discussed in Section~\ref{sec:doe:definefms} or automatically extracted using the method proposed in Section~\ref{sec:micomp}. FMs can be considered as functions, $f_1(.)$ to $f_q(.)$, which when applied to individual output observations, return scalar values. Taken column-wise, these values are IID (because they are obtained from IID replications), constituting a \textit{sample} prone to statistical analysis, or more specifically, to hypothesis testing. There will be as many samples (or groups of observations) per FM as there are model implementations. For example, considering some conceptual model with a single output $X$, FM samples for the $h^{\text{th}}$ implementation would be constituted as follows:

\begin{align*}
\mathbf{f}_1^h =& \begin{bmatrix} f_1(\mathbf{X}_1^h) & f_1(\mathbf{X}_2^h) & \ldots & f_1(\mathbf{X}_n^h) \end{bmatrix} \\
\mathbf{f}_2^h =& \begin{bmatrix} f_2(\mathbf{X}_1^h) & f_2(\mathbf{X}_2^h) & \ldots & f_2(\mathbf{X}_n^h) \end{bmatrix} \\
\vdots & \\
\mathbf{f}_q^h =& \begin{bmatrix} f_q(\mathbf{X}_1^h) & f_q(\mathbf{X}_2^h) & \ldots & f_q(\mathbf{X}_n^h) \end{bmatrix} \\
\end{align*}

Alternatively, FMs can be combined into one multivariate FM, e.g., for the $h^{\text{th}}$ implementation we would have:

\begin{equation*}
	\mathbf{F}^h=
	\begin{bmatrix}
		f_1(\mathbf{X}_1^h) & f_2(\mathbf{X}_1^h) & \ldots & f_q(\mathbf{X}_1^h) \\
		f_1(\mathbf{X}_2^h) & f_2(\mathbf{X}_2^h) & \ldots & f_q(\mathbf{X}_2^h) \\
		\vdots & \vdots & \ddots & \vdots \\
		f_1(\mathbf{X}_n^h) & f_2(\mathbf{X}_n^h) & \ldots & f_q(\mathbf{X}_n^h) \\
	\end{bmatrix}
\end{equation*}

\noindent where rows correspond to observations, and columns to variables or dimensions. 

\subsection{Distributional analysis of FMs and choice of statistical tests}
\label{sec:doe:definestatmeth}

In order to choose adequate tests for comparing each FM, it is necessary to first check if the FM samples are in accordance with the assumptions described in Section~\ref{sec:tests}. If an FM appears to follow the normal distribution and if the respective samples have similar variance, then the parametric tests listed in Table~\ref{tab:stattests} will be suitable. Conversely, non-parametric tests will be acceptable if such assumptions are not verified, or if the median is considered to better represent the FM in question.

If the choice falls on a multivariate test, the presentation of results becomes considerably simpler, since there is only one $p$-value to  exhibit. However, it is important to keep in mind that the distributional analysis of the combined multidimensional FM becomes significantly more complex. Thus, it may be difficult to assess the suitability of MANOVA, or even of the non-parametric alternatives, for comparing the multidimensional FM. Furthermore, when a significant difference exists, the multivariate test may mask where such difference occurs, while univariate tests will clearly show which FMs are affected.

\subsection{Compare FMs using hypothesis tests}
\label{sec:doe:comparefms}

In the univariate case, a statistical test is applied per FM to samples from the compared models. For example, considering two model implementations with three FMs, three tests would be performed. The first test on samples $\mathbf{f}_1^1$ and $\mathbf{f}_1^2$, the second on samples $\mathbf{f}_2^1$ and $\mathbf{f}_2^2$, and the third on samples $\mathbf{f}_3^1$ and $\mathbf{f}_3^2$. Each test returns a $p$-value, so we could perform a multiple comparisons correction or present the uncorrected $p$-values keeping in mind the caveat of increased type I error likelihood.

Continuing with the previous example, if FMs are instead combined into a multidimensional FM, then a multivariate test (e.g., MANOVA) would be performed on samples $\mathbf{F}^1$ and $\mathbf{F}^2$, yielding a single $p$-value.

\subsection{Analyze results and decide on the alignment or otherwise of the compared model implementations}
\label{sec:doe:decide}

Hypothesis tests results should be analyzed per parameterization, because model implementations may be locally aligned for some parameter sets, but not for others. Thus, each parameterization should be considered a separate experiment, i.e., it should not be considered part of a multiple comparisons scenario. If at least one FM is shown to be statistically different for a given parameterization, then the compared implementations can be considered globally misaligned.

If the $p$-value(s) for a specific parameterization are above the chosen significance level $\alpha$, the implementations can be considered locally aligned for that parameter set. When one or more $p$-values are significant in a multiple comparison scenario (for the same parameterization), they should be in the proportion predicted by $\alpha$. No $p$-values should remain significant after Bonferroni-type corrections are performed. In the multivariate approach, where all the FMs are merged into one multidimensional FM, the single $p$-value should not be significant.

It is important to understand that statistical tests may not provide a definitive answer. Only in the case of clear misalignments, with very significant $p$-values, can one reject $H_0$ with confidence. In many situations some $p$-values may appear to be borderline significant, and is up to the modeler to judge if the detected misalignment has practical significance.

\section{Model-independent selection of FMs}
\label{sec:micomp}

The empirical selection of FMs, as described in Section~\ref{sec:doe:definefms}, has a number of disadvantages. First, it relies on summary measures which are model-dependent and, probably, user-dependent. 
Furthermore, for different model parameters, the originally selected FMs may be of no use, as simulation output may change substantially. For example, the warm-up period for the steady-state statistical summaries can be quite different. Certain models and/or parameterizations might not even display a steady-state behavior. Finally, it might not be clear which FMs best capture the behavior of a model.

A model-independent approach to FM selection should work directly from simulation output, automatically selecting the features that best explain potential differences between the implementations being compared, thus avoiding the disadvantages of an empirical selection. Additionally, such a method should not depend on the distributional properties of simulation output, and should be directly applicable by modelers.

Our proposal consists of automatically extracting the most relevant information from simulation output using PCA. PCA is a widely used technique \cite{fachada2014spectrometric} which extracts the largest variance components from data by applying singular value decomposition to a mean-centered data matrix. In other words, PCA is able to convert simulation output into a set of linearly uncorrelated measures which can be analyzed in a consistent, model-independent fashion. This technique is especially relevant for stochastic simulation models, as it considers not only equilibrium but also dynamics over time \cite{wilensky2007making}. Procedure~\ref{proc:micomp} summarizes this approach, replacing steps \ref{proc:modcompexp:fepar:selfms} and \ref{proc:modcompexp:fepar:extfms} of Procedure~\ref{proc:modcompexp}.

\begin{proc}
\caption{Obtaining model-independent summary measures from one generic simulation output from $s$ model implementations.}
\label{proc:micomp}

\begin{enumerate}
	\item Group the $X_{ji}^h$'s from all replications row-wise in matrix $\mathbf{X}$ for each implementation, as follows: 

\begin{equation*}
\mathbf{X}=
\begin{bmatrix}
X_{10}^1 & X_{11}^1 & \ldots & X_{1,m-1}^1  & X_{1,m}^1\\
X_{20}^1 & X_{21}^1 & \ldots & X_{2,m-1}^1  & X_{2,m}^1\\
\vdots & \vdots & \ddots & \vdots & \vdots \\
X_{n0}^1 & X_{n1}^1 & \ldots & X_{n,m-1}^1  & X_{n,m}^1\\
\vdots & \vdots & \ddots & \vdots & \vdots \\
X_{10}^s & X_{11}^s & \ldots & X_{1,m-1}^s  & X_{1,m}^s\\
X_{20}^s & X_{21}^s & \ldots & X_{2,m-1}^s  & X_{2,m}^s\\
\vdots & \vdots & \ddots & \vdots & \vdots \\
X_{n0}^s & X_{n1}^s & \ldots & X_{n,m-1}^s  & X_{n,m}^s\\
\end{bmatrix}
\end{equation*}
	
	\item Determine matrix $\mathbf{X}_c$, which is the column mean-centered version of $\mathbf{X}$.
	\item Apply PCA to matrix $\mathbf{X}_c$, considering that rows (replications) correspond to observations and columns (iterations or time steps) to variables. This yields: 
	\begin{itemize}
		\item Matrix $\mathbf{T}$, containing the representation of the original data in the principal components (PCs) space.

\begin{equation*}
\mathbf{T}=
\begin{bmatrix}
T_{11}^1 & T_{12}^1 & \ldots & T_{1,u-1}^1  & T_{1,u}^1\\
T_{21}^1 & T_{22}^1 & \ldots & T_{2,u-1}^1  & T_{2,u}^1\\
\vdots & \vdots & \ddots & \vdots & \vdots \\
T_{n1}^1 & T_{n2}^1 & \ldots & T_{n,u-1}^1  & T_{n,u}^1\\
\vdots & \vdots & \ddots & \vdots & \vdots \\
T_{11}^s & T_{12}^s & \ldots & T_{1,u-1}^s  & T_{1,u}^s\\
T_{21}^s & T_{22}^s & \ldots & T_{2,u-1}^s  & T_{2,u}^s\\
\vdots & \vdots & \ddots & \vdots & \vdots \\
T_{n1}^s & T_{n2}^s & \ldots & T_{n,u-1}^s  & T_{n,u}^s\\
\end{bmatrix}
\end{equation*}		
		
		\item Vector $\boldsymbol{\lambda}$, containing the eigenvalues of the covariance matrix of $\mathbf{X}_c$ in descending order, each eigenvalue corresponding to the variance of the columns of $\mathbf{T}$.
		
\begin{equation*}
\boldsymbol{\lambda}=
\begin{bmatrix}
\lambda_1 & \lambda_2 & \ldots & \lambda_{u-1} & \lambda_{u}
\end{bmatrix}
\end{equation*}	

	\end{itemize}

\end{enumerate}

\end{proc}

Procedure~\ref{proc:micomp} accepts, on a per output basis, the collected data (Procedure~\ref{proc:modcompexp}, step~\ref{proc:modcompexp:fepar:nrep}) in the form of matrix $\mathbf{X}$, yielding matrix $\mathbf{T}$, which contains the representation of $\mathbf{X}_c$ (i.e., the column mean-centered version of $\mathbf{X}$) in the principal components (PCs) space, as well as vector $\boldsymbol{\lambda}$, containing the eigenvalues of the covariance matrix of $\mathbf{X}_c$. The columns of $\mathbf{T}$ correspond to PCs, and are orderer by decreasing variance, i.e., the first column corresponds to the first PC, and so on. Rows of $\mathbf{T}$ correspond to observations. The $k^{\text{th}}$ column of $\mathbf{T}$ contains $sn$ model\hyp{}independent observations for the $k^{\text{th}}$ PC, $n$ for each implementation. Thus, each PC corresponds to an FM. 
As in the case of empirically selected FMs, univariate or multivariate statistical tests can be used to check if samples from different implementations are drawn from the same distribution. However, both testing approaches will not prioritize dimensions, even though the first PCs are more important for characterizing model differences, as they explain more variance. This can be handled in the univariate case by prioritizing PCs according to their explained variance using the weighted Bonferroni procedure on the resulting $p$-values \cite{rosenthal1983ensemble}. For multivariate tests, dimensions/variables can be limited to the number of PCs that explain a prespecified amount of variance, although there is no prioritization within the selected dimensions.

The eigenvalues vector $\boldsymbol{\lambda}$ is also important for this process, for two reasons: 1) to select a number of PCs (i.e., FMs) to be considered for hypothesis testing, such that these explain a prespecified percentage of variance; 2) the alignment or otherwise of $s$ model implementations can be empirically assessed by analyzing how the explained variance is distributed along PCs. The percentage of variance explained by each PC can be obtained as shown in Eq.~\ref{eq:pcvar}.

\begin{equation}
\label{eq:pcvar}
S^2_k(\%)=\frac{\lambda_k}{\sum{\boldsymbol{\lambda}}}
\end{equation}

\noindent where $k$ identifies the $k^{\text{th}}$ PC, $\lambda_k$ is the eigenvalue associated with the $k^{\text{th}}$ PC, and $\sum{\boldsymbol{\lambda}}$ is the sum of all eigenvalues. If the variance is well distributed along many PCs, it is an indication that the compared implementations are aligned, at least for the output being analyzed. On the other hand, if most of the variance is explained by the first PCs, it can be an indication that at least one model implementation is misaligned. The rationale being that if all implementations show the same dynamical behavior, then the projection of their outputs in the PC space will be close together and have similar statistics, i.e., means, medians and variance. As such, PCA will be unable to find components which explain large quantities of variance, and the variance will be well distributed along the PCs. If at least one model implementation is misaligned, the projection of its outputs in the PC space will be farther apart than the projections of the remaining implementations. As such, PCA will yield at least one component which explains large part of the overall variance.

The alignment of two or more implementations can be assessed by analyzing the following information: 1) the $p$-values produced by the univariate and multivariate statistical tests, which should be above the typical 1\% or 5\% significance levels in case of implementation alignment; in the univariate case, it may be useful to adjust the $p$-values using the weighted Bonferroni procedure to account for multiple comparisons; 2) in case of misalignment, the total number of PCs required to explain a prespecified amount of variance should be lower than in case of alignment; also, more variance should be explained by the first PCs of the former than by the same PCs of the latter; and, 3) the scatter plot of the first two PC dimensions, which can offer visual, although subjective feedback on model alignment; e.g., in case of misalignment, points associated with runs from different implementations should form distinct groups.

\subsection{Extension to multiple outputs}
\label{sec:micomp:multout}

Determining the alignment of models with multiple outputs may be less straightforward. If model implementations are clearly aligned or misaligned, conclusions can be drawn by analyzing the information provided by the proposed method applied to each one of the model outputs. 

In order to compare all model outputs simultaneously, a Bonferroni or similar $p$-value correction could be used. This strategy can be directly applicable in the multivariate case, since there is one $p$-value per model output. However, a direct application to the univariate case would be more complex, since there will be multiple $p$-values per model output (one $p$-value per PC), and these may have been previously corrected with the weighted Bonferroni procedure according to their explained variance. 

We propose an alternative approach based on the concatenation of all model outputs, centered and scaled. This reduces a model with $g$ outputs to a model with one output, which can be processed with Procedure~\ref{proc:micomp}. In order to perform output concatenation, outputs are centered and scaled such that their domains are in the same order of magnitude, replication-wise. This can be performed using range scaling on each output, for example, as shown below for a given simulation output;

\begin{equation}
\widetilde{X}_{ji}=\frac{X_{ji}-\mean{\mathbf{X}}_j}{\max{\mathbf{X}_j} - \min{\mathbf{X}_j}}, \qquad i=0,1,\ldots,m;\ j=1,\ldots,sn
\label{eq:rangescale}
\end{equation}

\begin{equation}
\widetilde{\mathbf{X}}_j=\left[ \widetilde{X}_{j0} \quad \widetilde{X}_{j1} \quad \ldots \quad \widetilde{X}_{jm} \right]
\label{eq:scaledoutput}
\end{equation}

\noindent where $i$ represents iterations or time steps, $j$ is the replication number, $X_{ji}$ is the output value at iteration $i$ and replication $j$, $\mathbf{X}_j$ is a row vector with the complete output generated in the $j^{\text{th}}$ replication, and $\widetilde{\mathbf{X}}_j$ is its range scaled version. Other centering and scaling methods, such as auto-scaling or level scaling \cite{berg2006centering}, can be used as an alternative to range scaling in Eq. \ref{eq:rangescale}. For a model with $g$ outputs, the resulting concatenated output for the $j^{\text{th}}$ replication is given by

\begin{equation}
\widetilde{\mathbf{A}}_j={}_{1}\widetilde{\mathbf{X}}_j \oplus {}_{2}\widetilde{\mathbf{X}}_j \oplus \ldots \oplus {}_{g}\widetilde{\mathbf{X}}_j
\label{eq:concatoutput}
\end{equation}

\noindent where $\oplus$ is the concatenation operator, and $\widetilde{\mathbf{A}}_j$ is the concatenation of all model outputs for replication $j$. Model implementations can thus be compared with the proposed method using the ``single'' model output.

\section{Simulation model}
\label{sec:simmod}

The Predator\textendash{}Prey for High-Performance Computing (PPHPC) model is a reference model for studying and evaluating spatial ABM (SABM) implementation strategies, capturing important characteristics of SABMs, such as agent movement and local agent interactions. It is used in this work as a test case for the proposed model comparison method. The model is thoroughly described in reference \cite{fachada2015template} using the ODD protocol \cite{grimm2010odd}. Here we present a summarized description of the model.

\subsection{Description}
\label{sec:simmod:desc}

PPHPC is a predator\textendash{}prey model composed of three entity classes: \textit{agents}, \textit{grid cells} and \textit{environment}. \textit{Agents} can be of type prey or predator. While prey consume passive cell-bound food, predators consume prey. Agents have an energy state variable, $E$, which increases when feeding and decreases when moving and reproducing. When energy reaches zero, the agent is removed from the simulation. Instances of the \textit{grid cell} entity class are where agents act, namely where they try to feed and reproduce. Grid cells have a fixed grid position and contain only one resource, cell-bound food (\textit{grass}), which can be consumed by prey, and is represented by the countdown state variable $C$. The $C$ state variable specifies the number of iterations left for the cell-bound food to become available. Food becomes available when $C=0$, and when a prey consumes it, $C$ is set to $c_r$ (an initial simulation parameter).  The set of all grid cells forms the \textit{environment} entity, a toroidal square grid where the simulation takes place. The environment is defined by its size  and by the restart parameter, $c_r$. The temporal scale is represented by a positive integer $m$, which represents the number of iterations. 

Simulations start with an initialization process, where a predetermined number of agents are randomly placed in the simulation environment. Cell-bound food is also initialized at this stage. After initialization, and to get the simulation state at iteration zero, outputs are collected. The scheduler then enters the main simulation loop, where each iteration is sub-divided into four steps: 1) agent movement; 2) food growth in grid cells; 3) agent actions; and, 4) gathering of simulation outputs. Note that the following processes are explicitly random: a) initialization of specific state variables (e.g., initial agent energy); b) agent movement; c) the order in which agents act; and, d) agent reproduction. For process c), this implies that the agent list should be explicitly shuffled before agents can act.
 
Six outputs are collected at each iteration $i$: $P_i^s$, $P_i^w$, $P_i^c$, $\mean{E}_i^s$, $\mean{E}_i^w$, and $\mean{C}_i$. $P^s_i$ and $P^w_i$ refer to the total prey (\textit{sheep}) and predator (\textit{wolf}) population counts, respectively, while $P^c_i$ holds the quantity of available cell-bound food. $\mean{E}^s_i$ and $\mean{E}^w_i$ contain the mean energy of prey and predator populations. Finally, $\mean{C}_i$ refers to the mean value of the $C$ state variable in all grid cells.

\subsection{Parameterizations}
\label{sec:simmod:params}

Reference parameters for the PPHPC model are specified in reference \cite{fachada2015template}. Parameters are qualitatively separated into size-related and dynamics-related groups. Although size-related parameters also influence model dynamics, this separation is useful for parameterizing simulations.

Concerning size-related parameters, a base grid size of $100 \times 100$ is associated with $400$ prey and $200$ predators. Different grid sizes should have proportionally assigned agent population sizes, such that the initial agent density and the initial ratio between prey and predators remains constant. We define \textit{model size} as the association between grid size and initial agent population. For example, model size 200 corresponds to a grid size of $200 \times 200$ with \num{1600} prey and \num{800} predators at iteration 0.

For the dynamics-related parameters, two parameter sets, 1 and 2, are proposed. The two parameterizations generate distinct dynamics, with the second set typically yielding more than twice the number of agents than the first during the course of a simulation. We will refer to a combination of model size and parameter set as ``size$@$set'', e.g., $400@1$ for model size 400, parameter set 1. The reference number of iterations, $m$, is \num{4000}, not counting with the initial simulation state at iteration 0.

\subsection{Empirical selection of FMs}
\label{sec:simmod:compare}

Under the parameterizations discussed in the previous subsection, all outputs have an initial transient stage, entering steady-state after a number of iterations. The point at which simulations enter this state of equilibrium varies considerably from parameter set 1 to parameter set 2, occurring considerably sooner in the former. Model size does not seem to affect the steady-state truncation point, mainly influencing the magnitude of the collected outputs. As such, steady-state is empirically established for $i>1000$ and $i>2000$ for parameter sets 1 and 2, respectively \cite{fachada2015template}.

Considering this information, as well as the recommendations discussed in Section~\ref{sec:doe:definefms}, the following statistical summaries are selected for individual outputs: 1) maximum value ($\max$); 2) iteration where maximum value occurs ($\argmax$); 3) minimum value ($\min$); 4) iteration where minimum value occurs ($\argmin$); 5) steady-state mean ($\mean{X}^{\text{ss}}$); and, 6) steady-state sample standard deviation ($S^{\text{ss}}$). Thus, we specify a total of 36 FMs (six statistical summaries from six outputs). Naturally, the steady-state measures are collected after the truncation point defined for each parameter set.

As discussed in reference \cite{fachada2015template}, all measures, except $\argmax$ and $\argmin$, are amenable to be compared using parametric methods, as they follow (or approximately follow) normal distributions.

\subsection{Implementations}
\label{sec:simmod:impls}

Two implementations of the PPHPC model are used for evaluating the proposed model comparison technique. The first is developed in NetLogo \cite{fachada2015template}, while the second is a Java implementation with several parallel variants \cite{fachada2015parallelization}. For the results discussed in this paper, simulations performed with the Java implementation were executed with the EX parallelization strategy using eight threads. In this strategy, each thread processes an equal part of the simulation environment, and reproducible simulations are guaranteed. Individual threads use their own sub-sequence of a global random sequence, obtained with a random spacing approach using the SHA-256 cryptographic hash function. The Mersenne Twister pseudo-random number generator (PRNG) \cite{matsumoto1998mersenne} is used by both implementations for driving the model's random processes. Complete details of both implementations are available in the provided references, and their source code is available at \url{https://github.com/fakenmc/pphpc/}.

\section{Experimental setup}
\label{sec:expsetup}

In order to test the model comparison methods, we define a base PPHPC configuration using the NetLogo implementation and compare it with three configurations using the Java implementation. All configurations are tested for model sizes 100, 200, 400 and 800, and parameter sets 1 and 2, as described in Section \ref{sec:simmod:params}. The four configurations follow the conceptual model, except when stated otherwise:

\begin{enumerate}
\item NetLogo implementation.
\item Java implementation.
\item Java implementation: agents are ordered by energy prior to agent actions, and agent list shuffling is not performed.
\item Java implementation: the restart parameter, $c_r$, is set to one unit less than specified in the tested parameterizations (9 instead of 10 for parameter set 1, 14 instead of 15 for parameter set 2). 
\end{enumerate}

The goal is to assess how the two FM selection strategies (empirical and model-independent) expose the increasing differences of comparing configuration 1 with configurations 2\textendash{}4. More specifically, we are interested in checking if the proposed model-independent comparison method is able to expose these differences in the same way as the manual or empirical approach. As such, we define three comparison cases:

\begin{description}
\item[Case I] Compare configuration 1 with configuration 2. These configurations should yield distributionally equivalent results.
\item[Case II] Compare configuration 1 with configuration 3. A small misalignment is to be expected.
\item[Case III] Compare configuration 1 with configuration 4. There should be a mismatch in the outputs.
\end{description}

For each ``size$@$set'' combination, independent samples of the six model outputs are obtained from $n=30$ replications for each configuration, in a total of $4n=120$ runs. Each replication $r=1,\ldots,4n$ is performed with a PRNG seed obtained by taking the MD5 checksum of $r$ and converting the resulting hexadecimal string to an integer (limited to 32-bits for NetLogo and 128-bits for the Java implementation), guaranteeing independence between seeds, and consequently, between replications. The same samples are used for the evaluation of both empirical and model-independent FM selection.

In order to evaluate how the tested methodologies respond to larger sample sizes, an additional $n=100$ replications were performed per configuration for the $400@1$ combination, in a total of $4n=400$ runs. The PRNG seeds for individual replications were obtained in the same fashion. Similarly, the tested methodologies were also compared with $n=10$ for the $400@1$ combination, but in this case using the first 10 replications from the $n=30$ setup for each configuration. All samples sizes were chosen for convenience with the purpose of simplifying analysis of results, namely of how the two FM selection approaches, empirical and model-independent, fare under the same conditions.

For the model-dependent comparisons, results were obtained with the \swpackage{SimOutUtils} MATLAB toolbox \cite{fachada2016simoututils}. The \swpackage{micompr} R package \cite{fachada2016micompr} provides an implementation of the proposed model-independent comparison approach, and was used to produce the corresponding results.

The data generated by this computational experiment, as well as the scripts used to set up the experiment, are made available to other researchers at \url{https://zenodo.org/record/46848}.

\section{Results}
\label{sec:results}

In this section we mainly focus on the results for the $400@1$ combination. Results for the remaining size/set combinations are provided as Supplementary material, and are referred to when appropriate.

\begin{table}[!htbp]
\begin{center}
\begin{tabular}{clrrrrrr}
\toprule
\multirow{2}{*}{Comp.} & \multirow{2}{*}{Stat.} & \multicolumn{6}{c}{Outputs} \\
\cmidrule(l){3-8}
 & & $P^s$ & $P^w$ & $P^c$ & $\overline{E}^s$ & $\overline{E}^w$ & $\overline{C}$ \\
\midrule
\multirow{6}{*}{I} & 
$\max$ &  0.213 &  0.288 & \uline{ 0.011} &  0.231 &  0.774 &  0.086\\
& $\argmax$ &  0.733 &  0.802 &  0.056 &  0.858 &  0.284 &  0.279\\
& $\min$ & \uuline{ 0.002} &  0.088 &  0.094 &  0.275 &  0.076 & \uline{ 0.011}\\
& $\argmin$ & \uline{ 0.048} &  0.350 &  0.279 &  1.000 &  0.091 &  1.000\\
& $\mean{X}^{\text{ss}}$ &  0.457 &  0.905 &  0.546 &  0.833 & \uline{ 0.049} &  0.551\\
& $S^{\text{ss}}$ &  0.242 &  0.282 &  0.285 &  0.474 &  0.109 &  0.285\\
\midrule
\multirow{6}{*}{II} & 
$\max$ & \uuline{\num[output-exponent-marker=\text{e}]{7e-06}} & \uuline{\num[output-exponent-marker=\text{e}]{1e-07}} & \uuline{\num[output-exponent-marker=\text{e}]{2e-04}} & \uuline{\num[output-exponent-marker=\text{e}]{2e-07}} &  0.322 & \uuline{ 0.004}\\
& $\argmax$ & \uuline{\num[output-exponent-marker=\text{e}]{1e-04}} &  0.540 &  0.230 &  0.703 &  0.052 & \uuline{\num[output-exponent-marker=\text{e}]{1e-04}}\\
& $\min$ & \uuline{\num[output-exponent-marker=\text{e}]{1e-05}} & \uuline{\num[output-exponent-marker=\text{e}]{4e-04}} & \uuline{ 0.009} &  0.508 & \uline{ 0.017} & \uuline{ 0.001}\\
& $\argmin$ &  0.712 &  0.906 & \uuline{ 0.001} &  1.000 & \uuline{ 0.009} &  1.000\\
& $\mean{X}^{\text{ss}}$ &  0.206 & \uuline{< \num[output-exponent-marker=\text{e}]{1e-08}} & \uuline{\num[output-exponent-marker=\text{e}]{6e-08}} & \uuline{< \num[output-exponent-marker=\text{e}]{1e-08}} &  0.938 & \uuline{\num[output-exponent-marker=\text{e}]{6e-08}}\\
& $S^{\text{ss}}$ &  0.735 &  0.324 &  0.712 &  0.218 &  0.688 &  0.713\\
\midrule
\multirow{6}{*}{III} & 
$\max$ & \uuline{< \num[output-exponent-marker=\text{e}]{1e-08}} & \uuline{< \num[output-exponent-marker=\text{e}]{1e-08}} & \uuline{< \num[output-exponent-marker=\text{e}]{1e-08}} & \uuline{< \num[output-exponent-marker=\text{e}]{1e-08}} & \uuline{< \num[output-exponent-marker=\text{e}]{1e-08}} & \uuline{< \num[output-exponent-marker=\text{e}]{1e-08}}\\
& $\argmax$ & \uuline{\num[output-exponent-marker=\text{e}]{2e-05}} &  0.115 & \uuline{< \num[output-exponent-marker=\text{e}]{1e-08}} &  0.134 &  0.332 & \uuline{\num[output-exponent-marker=\text{e}]{1e-05}}\\
& $\min$ & \uuline{\num[output-exponent-marker=\text{e}]{7e-08}} & \uuline{< \num[output-exponent-marker=\text{e}]{1e-08}} & \uuline{< \num[output-exponent-marker=\text{e}]{1e-08}} &  0.582 & \uuline{ 0.003} & \uuline{< \num[output-exponent-marker=\text{e}]{1e-08}}\\
& $\argmin$ & \uuline{\num[output-exponent-marker=\text{e}]{2e-05}} &  0.070 & \uuline{\num[output-exponent-marker=\text{e}]{2e-05}} &  1.000 & \uuline{ 0.004} & \uuline{< \num[output-exponent-marker=\text{e}]{1e-08}}\\
& $\mean{X}^{\text{ss}}$ & \uuline{< \num[output-exponent-marker=\text{e}]{1e-08}} & \uuline{< \num[output-exponent-marker=\text{e}]{1e-08}} & \uuline{< \num[output-exponent-marker=\text{e}]{1e-08}} & \uuline{< \num[output-exponent-marker=\text{e}]{1e-08}} & \uuline{< \num[output-exponent-marker=\text{e}]{1e-08}} & \uuline{< \num[output-exponent-marker=\text{e}]{1e-08}}\\
& $S^{\text{ss}}$ & \uuline{\num[output-exponent-marker=\text{e}]{3e-07}} & \uuline{\num[output-exponent-marker=\text{e}]{7e-08}} & \uuline{\num[output-exponent-marker=\text{e}]{5e-05}} & \uuline{\num[output-exponent-marker=\text{e}]{4e-08}} & \uline{ 0.013} & \uline{ 0.013}\\
\bottomrule
\end{tabular}

\caption{\label{tab:mdcomps}$P$-values for the empirical selection of FMs for model size 400, parameter set 1, and $n=30$ runs per configuration. $P$-values were obtained with the $t$-test for $\max$, $\min$, $\mean{X}^{\text{ss}}$ and $S^{\text{ss}}$ statistics, and with the Mann-Whitney U test for $\argmax$ and $\argmin$ statistics. Case I weights two similar configurations, case II compares configurations with a small implementation difference, and case III compares configurations with a different parameter. Values lower than $0.05$ are underlined, while values lower than $0.01$ are double-underlined.}
\end{center}
\end{table}

\subsection{Empirical selection of FMs}
\label{sec:results:empfms}

Table~\ref{tab:mdcomps} shows the results for the empirical approach to cases I, II and III. The corresponding $p$-values were obtained with the $t$-test for $\max$, $\min$, $\mean{X}^{\text{ss}}$ and $S^{\text{ss}}$ statistics, and with the Mann\textendash{}Whitney \textit{U} test for the $\argmax$ and $\argmin$ statistics. For the latter, when all points in both samples have the same value, we simply present 1.00 as the $p$-value, since the test is not performed. 

Results for case I show that the two configurations are reasonably well aligned. In a total of 36 FMs, five are below the 5\% significance level, and of these, only one is below 1\%. Since the $p$-values are not corrected for multiple comparisons, it is expected that some of them present ``significant'' differences. For case II, the effect of not shuffling the agents before they act in the Java implementation (configuration 3) is evident, as half of the $p$-values are below 0.05. Of these, all but one are also below 0.01. Finally, in case III, most $p$-values show very significant differences, such that we can reject, with a high degree of confidence, the hypothesis that the corresponding FMs are produced by equivalent configurations.

Tables~S1.1\textendash{}S1.8, provided as Supplementary material, show the results for the remaining size/set combinations. The general tendency is the same, i.e., $p$-values for case I highlight very few to no significant differences between the tested configurations (although there are, sporadicly, some significant differences in the $\mean{E}^s$ output), while for cases II and III, the dissimilarities are apparent. However, perhaps unexpectedly, results for parameter set 2 indicate that the configurations compared in case II have more significant differences than those in case III. 

\begin{table}[!htbp]
\begin{center}
\resizebox{\columnwidth}{!}{%
\begin{tabular}{clrrrrrrr}
\toprule
\multirow{2}{*}{Comp.} & \multirow{2}{*}{Data} & \multicolumn{7}{c}{Outputs} \\
\cmidrule(l){3-9}
 &  & $P^s$ & $P^w$ & $P^c$ & $\mean{E}^s$ & $\mean{E}^w$ & $\mean{C}$ & $\widetilde{A}$\\
\midrule
\multirow{4}{*}{I}
 & $\#$PCs & 24 & 32 & 25 & 36 & 43 & 26 & 39\\
 & MNV & 0.528 & 0.258 & 0.548 & 0.105 & 0.746 & 0.577 & 0.704\\
 & $t$-test & 0.530 & 0.836 & 0.804 & 0.784 & 0.378 & 0.805 & 0.879\\
 & PCS & \raisebox{-.5\height}{\resizebox {1.2cm} {1.2cm} { \begin{tikzpicture}[scale=6] \path (-1.2,-1.2) (1.2,1.2);\draw[very thin,color=gray] (0,1.1)--(0,-1.1); \draw[very thin,color=gray] (1.1,0)--(-1.1,0); \path plot[mark=square*,mark options={color=red},mark size=0.8pt] coordinates { (-0.664,0.010) (-0.801,0.094) (0.313,0.051) (-0.144,-0.250) (0.223,0.241) (0.211,-0.246) (-0.333,-0.005) (0.956,0.258) (0.280,0.114) (-0.150,-0.331) (0.515,0.091) (0.030,-0.002) (-0.151,-0.034) (-0.379,0.703) (0.237,-0.165) (-0.224,0.151) (0.338,0.258) (0.594,0.703) (-0.275,0.125) (0.241,-0.208) (0.506,-0.037) (0.921,-0.306) (-0.053,-0.061) (0.359,0.074) (-1.000,0.352) (-0.057,0.068) (-0.153,-0.554) (-0.287,0.239) (0.076,0.330) (-0.130,0.750)};  \path plot[mark=diamond*,mark size=1pt] coordinates { (0.146,0.204) (0.156,-0.377) (-0.670,-0.205) (0.516,0.020) (-0.095,-0.147) (0.461,0.112) (-0.533,0.207) (-0.521,0.078) (0.060,-0.341) (0.434,-0.115) (0.252,-0.108) (-0.108,-0.193) (-0.782,-0.153) (0.086,-0.250) (0.193,-0.338) (-0.137,-0.026) (-0.176,-0.141) (0.210,0.023) (0.416,0.476) (-0.326,0.142) (-0.188,-0.274) (-0.366,-0.086) (-0.227,-0.059) (-0.017,-0.345) (0.587,-0.370) (-0.279,-0.204) (0.198,-0.052) (-0.313,0.138) (0.313,0.292) (-0.288,-0.326)};  \end{tikzpicture} }} & \raisebox{-.5\height}{\resizebox {1.2cm} {1.2cm} { \begin{tikzpicture}[scale=6] \path (-1.2,-1.2) (1.2,1.2);\draw[very thin,color=gray] (0,1.1)--(0,-1.1); \draw[very thin,color=gray] (1.1,0)--(-1.1,0); \path plot[mark=square*,mark options={color=red},mark size=0.8pt] coordinates { (-0.632,0.161) (-0.778,0.155) (0.223,-0.213) (-0.072,0.179) (0.150,-0.120) (0.230,0.143) (-0.315,-0.049) (0.869,-0.226) (0.216,-0.019) (-0.082,0.559) (0.459,-0.179) (0.085,-0.062) (-0.093,0.099) (-0.546,-0.591) (0.328,0.052) (-0.321,-0.209) (0.170,-0.394) (0.404,-0.664) (-0.374,-0.194) (0.282,0.149) (0.354,-0.090) (1.000,0.167) (0.151,0.212) (0.183,-0.269) (-0.943,-0.221) (-0.130,0.092) (0.116,0.666) (-0.262,-0.253) (-0.053,-0.311) (-0.310,-0.698)};  \path plot[mark=diamond*,mark size=1pt] coordinates { (0.075,-0.182) (0.220,0.285) (-0.529,0.355) (0.492,-0.039) (-0.095,0.322) (0.307,-0.195) (-0.528,-0.128) (-0.492,-0.068) (0.077,0.268) (0.406,-0.042) (0.324,0.032) (-0.091,0.209) (-0.719,0.205) (0.126,0.347) (0.276,0.344) (0.025,0.135) (-0.084,0.198) (0.209,-0.169) (0.233,-0.535) (-0.327,-0.319) (-0.057,0.242) (-0.338,0.202) (-0.112,0.016) (-0.003,0.311) (0.589,0.187) (-0.198,0.207) (0.288,0.144) (-0.378,-0.162) (0.149,-0.478) (-0.154,0.437)};  \end{tikzpicture} }} & \raisebox{-.5\height}{\resizebox {1.2cm} {1.2cm} { \begin{tikzpicture}[scale=6] \path (-1.2,-1.2) (1.2,1.2);\draw[very thin,color=gray] (0,1.1)--(0,-1.1); \draw[very thin,color=gray] (1.1,0)--(-1.1,0); \path plot[mark=square*,mark options={color=red},mark size=0.8pt] coordinates { (0.613,-0.071) (0.723,0.021) (-0.219,0.042) (0.073,-0.256) (-0.138,0.273) (-0.217,-0.239) (0.300,-0.069) (-0.831,0.432) (-0.248,0.171) (0.017,-0.289) (-0.462,0.182) (-0.034,-0.006) (0.082,-0.053) (0.501,0.700) (-0.241,-0.144) (0.235,0.098) (-0.226,0.258) (-0.374,0.814) (0.333,0.060) (-0.270,-0.168) (-0.464,0.025) (-0.895,-0.201) (-0.004,-0.058) (-0.282,0.108) (1.000,0.225) (0.066,0.129) (-0.047,-0.537) (0.308,0.262) (0.015,0.298) (0.321,0.673)};  \path plot[mark=diamond*,mark size=1pt] coordinates { (-0.101,0.242) (-0.222,-0.356) (0.546,-0.271) (-0.501,0.088) (0.054,-0.107) (-0.423,0.218) (0.545,0.178) (0.536,-0.048) (-0.115,-0.387) (-0.446,-0.090) (-0.222,-0.121) (0.087,-0.271) (0.697,-0.250) (-0.152,-0.244) (-0.269,-0.305) (0.116,-0.045) (0.122,-0.174) (-0.174,-0.001) (-0.260,0.533) (0.333,0.067) (0.083,-0.275) (0.271,-0.070) (0.186,-0.015) (-0.075,-0.357) (-0.615,-0.325) (0.228,-0.265) (-0.223,0.022) (0.363,0.069) (-0.189,0.255) (0.188,-0.375)};  \end{tikzpicture} }} & \raisebox{-.5\height}{\resizebox {1.2cm} {1.2cm} { \begin{tikzpicture}[scale=6] \path (-1.2,-1.2) (1.2,1.2);\draw[very thin,color=gray] (0,1.1)--(0,-1.1); \draw[very thin,color=gray] (1.1,0)--(-1.1,0); \path plot[mark=square*,mark options={color=red},mark size=0.8pt] coordinates { (0.760,-0.049) (0.625,0.290) (-0.263,-0.272) (0.187,-0.380) (-0.314,0.409) (-0.069,-0.474) (0.459,-0.095) (-0.596,0.306) (-0.322,0.214) (0.055,-0.035) (-0.302,0.062) (-0.053,-0.004) (0.322,-0.038) (0.458,1.000) (-0.085,-0.455) (0.086,0.150) (-0.325,0.110) (-0.327,0.476) (0.336,-0.119) (-0.160,-0.281) (-0.611,0.045) (-0.882,-0.050) (0.174,-0.055) (-0.250,-0.404) (0.954,0.215) (0.059,0.003) (0.220,-0.172) (0.189,0.332) (-0.017,0.322) (0.101,0.662)};  \path plot[mark=diamond*,mark size=1pt] coordinates { (-0.101,0.215) (-0.167,-0.236) (0.630,-0.236) (-0.618,0.345) (-0.120,0.067) (-0.318,-0.014) (0.409,0.041) (0.497,-0.177) (-0.106,-0.312) (-0.269,0.039) (-0.199,-0.160) (0.078,-0.335) (0.532,-0.066) (-0.242,-0.061) (-0.493,-0.220) (0.125,0.085) (0.001,-0.101) (-0.117,-0.046) (-0.541,0.489) (0.352,0.020) (0.062,-0.121) (0.335,0.151) (0.278,-0.028) (-0.023,-0.207) (-0.821,-0.352) (0.243,-0.330) (0.043,0.093) (0.241,-0.233) (-0.290,0.190) (0.193,-0.215)};  \end{tikzpicture} }} & \raisebox{-.5\height}{\resizebox {1.2cm} {1.2cm} { \begin{tikzpicture}[scale=6] \path (-1.2,-1.2) (1.2,1.2);\draw[very thin,color=gray] (0,1.1)--(0,-1.1); \draw[very thin,color=gray] (1.1,0)--(-1.1,0); \path plot[mark=square*,mark options={color=red},mark size=0.8pt] coordinates { (0.239,-0.084) (-0.329,0.400) (-0.040,0.354) (0.252,0.019) (0.166,-0.123) (0.322,0.185) (0.422,-0.103) (0.186,-0.133) (0.275,0.279) (0.367,-0.412) (-0.236,0.427) (-0.042,-0.484) (0.129,-0.160) (-0.145,-0.732) (0.279,0.199) (-0.128,0.225) (-0.199,0.168) (-0.121,0.295) (-0.427,-0.098) (-0.092,-0.254) (0.194,-0.154) (0.414,0.588) (0.532,0.075) (0.331,0.297) (-0.026,-0.045) (-0.212,0.045) (-0.373,-0.050) (-0.734,-0.474) (-0.073,-0.045) (0.222,-0.081)};  \path plot[mark=diamond*,mark size=1pt] coordinates { (-0.511,0.358) (0.764,0.162) (0.216,-0.155) (0.403,-0.573) (0.121,-0.225) (-0.250,0.026) (-0.335,-0.047) (1.000,-0.022) (-0.278,0.190) (0.155,-0.025) (0.048,0.022) (-0.052,0.455) (-0.073,0.218) (-0.403,-0.195) (0.099,-0.104) (0.302,0.090) (-0.386,-0.207) (-0.194,0.025) (-0.062,0.045) (-0.366,-0.046) (-0.251,0.132) (-0.670,-0.333) (-0.140,-0.158) (-0.532,0.486) (-0.206,0.606) (-0.077,-0.793) (0.073,-0.075) (0.404,-0.188) (0.165,0.118) (-0.115,0.093)};  \end{tikzpicture} }} & \raisebox{-.5\height}{\resizebox {1.2cm} {1.2cm} { \begin{tikzpicture}[scale=6] \path (-1.2,-1.2) (1.2,1.2);\draw[very thin,color=gray] (0,1.1)--(0,-1.1); \draw[very thin,color=gray] (1.1,0)--(-1.1,0); \path plot[mark=square*,mark options={color=red},mark size=0.8pt] coordinates { (0.613,-0.071) (0.723,0.021) (-0.219,0.042) (0.072,-0.256) (-0.138,0.273) (-0.217,-0.239) (0.300,-0.070) (-0.830,0.433) (-0.248,0.172) (0.017,-0.289) (-0.462,0.183) (-0.035,-0.007) (0.082,-0.053) (0.502,0.700) (-0.242,-0.144) (0.235,0.097) (-0.226,0.258) (-0.374,0.815) (0.333,0.059) (-0.271,-0.168) (-0.464,0.025) (-0.895,-0.200) (-0.004,-0.059) (-0.283,0.108) (1.000,0.225) (0.066,0.130) (-0.047,-0.538) (0.309,0.263) (0.015,0.298) (0.322,0.673)};  \path plot[mark=diamond*,mark size=1pt] coordinates { (-0.101,0.242) (-0.222,-0.356) (0.546,-0.272) (-0.501,0.088) (0.055,-0.106) (-0.423,0.218) (0.546,0.178) (0.536,-0.048) (-0.115,-0.388) (-0.446,-0.091) (-0.222,-0.121) (0.087,-0.271) (0.697,-0.251) (-0.152,-0.244) (-0.268,-0.305) (0.116,-0.045) (0.122,-0.174) (-0.174,-0.001) (-0.259,0.533) (0.333,0.067) (0.083,-0.275) (0.271,-0.070) (0.185,-0.015) (-0.075,-0.357) (-0.615,-0.325) (0.227,-0.265) (-0.224,0.022) (0.363,0.068) (-0.189,0.255) (0.188,-0.375)};  \end{tikzpicture} }} & \raisebox{-.5\height}{\resizebox {1.2cm} {1.2cm} { \begin{tikzpicture}[scale=6] \path (-1.2,-1.2) (1.2,1.2);\draw[very thin,color=gray] (0,1.1)--(0,-1.1); \draw[very thin,color=gray] (1.1,0)--(-1.1,0); \path plot[mark=square*,mark options={color=red},mark size=0.8pt] coordinates { (0.609,-0.155) (0.718,-0.115) (-0.195,0.125) (0.083,-0.274) (-0.138,0.188) (-0.213,-0.228) (0.319,-0.025) (-0.792,0.417) (-0.217,0.127) (0.004,-0.460) (-0.455,0.245) (-0.041,0.036) (0.073,-0.082) (0.654,0.628) (-0.302,-0.096) (0.267,0.185) (-0.129,0.368) (-0.230,0.646) (0.401,0.117) (-0.336,-0.115) (-0.402,0.138) (-0.901,-0.176) (-0.074,-0.198) (-0.176,0.181) (1.000,0.127) (0.083,-0.032) (-0.203,-0.588) (0.322,0.288) (0.091,0.313) (0.404,0.691)};  \path plot[mark=diamond*,mark size=1pt] coordinates { (-0.078,0.211) (-0.247,-0.320) (0.498,-0.389) (-0.482,0.164) (0.039,-0.262) (-0.319,0.229) (0.519,0.112) (0.542,-0.122) (-0.131,-0.295) (-0.427,0.097) (-0.310,-0.021) (0.074,-0.321) (0.643,-0.287) (-0.189,-0.246) (-0.357,-0.295) (0.025,-0.172) (0.068,-0.215) (-0.193,0.150) (-0.169,0.580) (0.342,0.225) (-0.026,-0.229) (0.246,-0.090) (0.139,-0.072) (-0.111,-0.293) (-0.675,-0.176) (0.185,-0.269) (-0.235,-0.057) (0.391,0.038) (-0.107,0.472) (0.119,-0.423)};  \end{tikzpicture} }}\\
\midrule
\multirow{4}{*}{II}
 & $\#$PCs & 24 & 29 & 24 & 12 & 43 & 25 & 38\\
 & MNV & \uuline{<1e-08} & \uuline{<1e-08} & \uuline{5e-08} & \uuline{<1e-08} & 0.467 & \uuline{1e-07} & \uuline{<1e-08}\\
 & $t$-test & \uline{0.042} & \uuline{<1e-08} & 0.108 & \uuline{<1e-08} & \uline{0.017} & 0.109 & 0.390\\
 & PCS & \raisebox{-.5\height}{\resizebox {1.2cm} {1.2cm} { \begin{tikzpicture}[scale=6] \path (-1.2,-1.2) (1.2,1.2);\draw[very thin,color=gray] (0,1.1)--(0,-1.1); \draw[very thin,color=gray] (1.1,0)--(-1.1,0); \path plot[mark=square*,mark options={color=red},mark size=0.8pt] coordinates { (-0.522,0.202) (-0.668,0.179) (0.418,0.063) (-0.061,-0.241) (0.266,0.131) (0.333,-0.202) (-0.276,-0.078) (1.000,0.245) (0.265,0.086) (-0.126,-0.121) (0.570,-0.025) (0.100,-0.048) (-0.144,-0.099) (-0.228,0.653) (0.245,-0.243) (-0.164,0.091) (0.492,0.286) (0.656,0.253) (-0.112,0.214) (0.278,-0.190) (0.571,-0.093) (0.935,-0.366) (-0.087,-0.180) (0.517,0.037) (-0.834,0.377) (0.038,0.063) (-0.171,-0.519) (-0.222,0.288) (0.169,0.207) (0.054,0.846)};  \path plot[mark=diamond*,mark size=1pt] coordinates { (0.098,0.111) (-0.030,-0.590) (0.006,0.333) (0.494,0.509) (-0.423,0.314) (-0.610,-0.485) (-0.187,0.114) (-0.840,-0.384) (-0.019,-0.225) (-0.125,-0.276) (-0.231,0.591) (-0.876,0.131) (-0.256,-0.061) (-0.384,-0.310) (0.647,-0.150) (-0.220,0.330) (-0.356,-0.103) (0.244,-0.227) (-0.073,-0.179) (-0.709,-0.407) (0.097,-0.018) (-0.096,-0.332) (-0.368,0.340) (0.561,-0.334) (-0.295,-0.140) (0.057,-0.432) (-0.078,-0.104) (0.054,-0.194) (0.535,-0.159) (0.095,0.526)};  \end{tikzpicture} }} & \raisebox{-.5\height}{\resizebox {1.2cm} {1.2cm} { \begin{tikzpicture}[scale=6] \path (-1.2,-1.2) (1.2,1.2);\draw[very thin,color=gray] (0,1.1)--(0,-1.1); \draw[very thin,color=gray] (1.1,0)--(-1.1,0); \path plot[mark=square*,mark options={color=red},mark size=0.8pt] coordinates { (-0.066,0.635) (-0.033,0.650) (-0.634,-0.113) (-0.417,0.085) (-0.598,0.057) (-0.473,-0.276) (-0.241,0.309) (-1.000,-0.412) (-0.577,0.105) (-0.182,0.100) (-0.732,-0.251) (-0.508,0.033) (-0.375,0.220) (-0.356,0.750) (-0.545,-0.137) (-0.336,0.452) (-0.705,-0.004) (-0.804,-0.119) (-0.343,0.404) (-0.494,-0.174) (-0.610,-0.316) (-0.959,-0.764) (-0.450,0.020) (-0.675,-0.168) (-0.127,0.999) (-0.309,0.098) (-0.273,-0.141) (-0.478,0.535) (-0.493,0.165) (-0.507,0.534)};  \path plot[mark=diamond*,mark size=1pt] coordinates { (0.424,-0.108) (0.476,-0.633) (0.293,0.064) (0.042,-0.044) (0.603,0.451) (0.918,-0.097) (0.560,0.062) (0.815,0.270) (0.510,-0.297) (0.516,-0.272) (0.576,0.415) (0.888,0.561) (0.496,0.052) (0.619,-0.075) (0.149,-0.741) (0.542,0.289) (0.588,0.104) (0.445,-0.633) (0.520,-0.213) (0.797,0.043) (0.355,-0.263) (0.441,-0.337) (0.401,0.432) (0.207,-0.731) (0.576,-0.032) (0.335,-0.512) (0.454,-0.194) (0.461,-0.367) (0.107,-0.688) (0.188,0.218)};  \end{tikzpicture} }} & \raisebox{-.5\height}{\resizebox {1.2cm} {1.2cm} { \begin{tikzpicture}[scale=6] \path (-1.2,-1.2) (1.2,1.2);\draw[very thin,color=gray] (0,1.1)--(0,-1.1); \draw[very thin,color=gray] (1.1,0)--(-1.1,0); \path plot[mark=square*,mark options={color=red},mark size=0.8pt] coordinates { (0.530,-0.165) (0.657,-0.087) (-0.352,-0.131) (-0.000,0.243) (-0.201,-0.162) (-0.368,0.174) (0.278,0.152) (-0.924,-0.409) (-0.204,-0.083) (0.033,0.154) (-0.567,-0.047) (-0.118,0.037) (0.125,0.171) (0.389,-0.660) (-0.261,0.242) (0.220,-0.031) (-0.424,-0.374) (-0.496,-0.334) (0.180,-0.245) (-0.334,0.160) (-0.605,0.010) (-1.000,0.290) (0.075,0.246) (-0.502,-0.126) (0.929,-0.285) (-0.039,-0.116) (0.036,0.655) (0.300,-0.291) (-0.089,-0.220) (0.150,-0.905)};  \path plot[mark=diamond*,mark size=1pt] coordinates { (-0.103,-0.167) (-0.088,0.639) (0.074,-0.392) (-0.423,-0.653) (0.503,-0.279) (0.452,0.595) (0.209,-0.112) (0.722,0.530) (-0.085,0.198) (0.048,0.318) (0.310,-0.648) (0.866,-0.032) (0.241,0.104) (0.263,0.375) (-0.667,0.056) (0.295,-0.363) (0.273,0.158) (-0.346,0.182) (0.033,0.171) (0.624,0.557) (-0.131,-0.005) (0.079,0.434) (0.424,-0.303) (-0.615,0.260) (0.298,0.183) (-0.145,0.479) (0.090,0.137) (-0.121,0.190) (-0.574,0.099) (0.077,-0.571)};  \end{tikzpicture} }} & \raisebox{-.5\height}{\resizebox {1.2cm} {1.2cm} { \begin{tikzpicture}[scale=6] \path (-1.2,-1.2) (1.2,1.2);\draw[very thin,color=gray] (0,1.1)--(0,-1.1); \draw[very thin,color=gray] (1.1,0)--(-1.1,0); \path plot[mark=square*,mark options={color=red},mark size=0.8pt] coordinates { (-0.877,0.287) (-0.960,0.248) (-0.815,-0.141) (-0.824,0.011) (-0.763,-0.079) (-0.784,-0.112) (-0.885,0.154) (-0.722,-0.229) (-0.861,-0.121) (-0.850,0.063) (-0.791,-0.134) (-0.871,-0.066) (-0.914,0.100) (-0.834,0.222) (-0.811,-0.073) (-0.841,0.039) (-0.833,-0.150) (-0.769,-0.145) (-0.893,0.071) (-0.809,-0.102) (-0.794,-0.230) (-0.700,-0.337) (-0.796,0.084) (-0.758,-0.160) (-0.883,0.353) (-0.853,0.012) (-1.000,0.035) (-0.839,0.094) (-0.784,0.005) (-0.828,0.086)};  \path plot[mark=diamond*,mark size=1pt] coordinates { (0.783,-0.079) (0.815,-0.132) (0.858,0.054) (0.913,-0.129) (0.851,0.179) (0.832,0.122) (0.861,0.019) (0.764,0.287) (0.928,-0.055) (0.924,0.011) (0.766,0.154) (0.725,0.346) (0.754,-0.017) (0.766,0.091) (0.887,-0.192) (0.866,0.095) (0.763,0.118) (0.867,-0.104) (0.842,-0.063) (0.678,0.177) (0.868,-0.095) (0.760,-0.032) (0.819,0.149) (0.876,-0.298) (0.834,0.046) (0.854,-0.092) (0.872,0.026) (0.834,-0.063) (0.897,-0.280) (0.886,-0.024)};  \end{tikzpicture} }} & \raisebox{-.5\height}{\resizebox {1.2cm} {1.2cm} { \begin{tikzpicture}[scale=6] \path (-1.2,-1.2) (1.2,1.2);\draw[very thin,color=gray] (0,1.1)--(0,-1.1); \draw[very thin,color=gray] (1.1,0)--(-1.1,0); \path plot[mark=square*,mark options={color=red},mark size=0.8pt] coordinates { (0.177,-0.178) (-0.523,0.086) (-0.287,0.380) (0.090,0.092) (0.054,0.121) (0.099,0.143) (0.205,-0.088) (0.049,0.029) (0.034,0.227) (0.395,-0.021) (-0.429,0.368) (0.050,-0.307) (-0.060,-0.003) (-0.054,-0.570) (0.076,0.395) (-0.333,0.095) (-0.274,-0.003) (-0.311,0.121) (-0.464,-0.286) (-0.127,0.097) (0.052,0.021) (0.089,0.774) (0.274,0.266) (0.104,0.220) (-0.204,-0.098) (-0.267,0.013) (-0.264,0.227) (-0.650,-0.435) (-0.263,-0.069) (0.051,-0.252)};  \path plot[mark=diamond*,mark size=1pt] coordinates { (-0.269,-0.025) (0.413,0.376) (0.489,-0.280) (-0.119,-0.487) (0.299,-0.262) (1.000,-0.221) (0.004,-0.048) (-0.273,0.059) (0.147,0.199) (0.140,0.101) (-0.501,-0.360) (-0.095,-0.120) (0.095,0.101) (0.263,0.037) (0.738,-0.160) (-0.247,-0.150) (0.230,0.079) (-0.029,0.366) (0.191,-0.314) (-0.324,0.045) (-0.042,-0.192) (0.128,-0.029) (0.301,-0.496) (-0.070,0.323) (0.250,0.131) (0.062,0.072) (-0.102,0.169) (0.053,0.218) (0.166,-0.015) (-0.187,-0.481)};  \end{tikzpicture} }} & \raisebox{-.5\height}{\resizebox {1.2cm} {1.2cm} { \begin{tikzpicture}[scale=6] \path (-1.2,-1.2) (1.2,1.2);\draw[very thin,color=gray] (0,1.1)--(0,-1.1); \draw[very thin,color=gray] (1.1,0)--(-1.1,0); \path plot[mark=square*,mark options={color=red},mark size=0.8pt] coordinates { (0.530,-0.164) (0.657,-0.086) (-0.352,-0.132) (-0.001,0.244) (-0.201,-0.163) (-0.369,0.173) (0.277,0.152) (-0.923,-0.411) (-0.205,-0.082) (0.034,0.153) (-0.567,-0.047) (-0.118,0.037) (0.125,0.172) (0.390,-0.660) (-0.262,0.243) (0.220,-0.029) (-0.424,-0.374) (-0.496,-0.334) (0.181,-0.245) (-0.335,0.160) (-0.605,0.009) (-1.000,0.288) (0.075,0.247) (-0.502,-0.126) (0.928,-0.283) (-0.038,-0.117) (0.034,0.655) (0.301,-0.291) (-0.089,-0.220) (0.152,-0.905)};  \path plot[mark=diamond*,mark size=1pt] coordinates { (-0.102,-0.169) (-0.089,0.639) (0.074,-0.392) (-0.421,-0.654) (0.504,-0.280) (0.452,0.595) (0.210,-0.112) (0.720,0.532) (-0.085,0.197) (0.047,0.318) (0.312,-0.649) (0.866,-0.032) (0.241,0.105) (0.262,0.376) (-0.667,0.055) (0.297,-0.364) (0.272,0.159) (-0.346,0.180) (0.033,0.171) (0.623,0.559) (-0.131,-0.005) (0.077,0.435) (0.424,-0.303) (-0.615,0.259) (0.298,0.183) (-0.146,0.479) (0.090,0.138) (-0.121,0.190) (-0.574,0.098) (0.077,-0.570)};  \end{tikzpicture} }} & \raisebox{-.5\height}{\resizebox {1.2cm} {1.2cm} { \begin{tikzpicture}[scale=6] \path (-1.2,-1.2) (1.2,1.2);\draw[very thin,color=gray] (0,1.1)--(0,-1.1); \draw[very thin,color=gray] (1.1,0)--(-1.1,0); \path plot[mark=square*,mark options={color=red},mark size=0.8pt] coordinates { (0.601,-0.084) (0.668,0.035) (-0.331,-0.136) (-0.006,0.243) (-0.167,-0.135) (-0.389,0.054) (0.266,0.173) (-0.807,-0.606) (-0.125,-0.125) (0.045,0.114) (-0.564,-0.224) (-0.127,-0.006) (0.100,0.214) (0.646,-0.542) (-0.322,0.251) (0.256,0.018) (-0.279,-0.493) (-0.317,-0.261) (0.299,-0.263) (-0.395,0.056) (-0.566,-0.180) (-1.000,0.115) (-0.002,0.443) (-0.416,-0.245) (0.991,-0.013) (-0.019,-0.109) (-0.188,0.655) (0.389,-0.272) (0.002,-0.222) (0.336,-0.919)};  \path plot[mark=diamond*,mark size=1pt] coordinates { (-0.064,-0.297) (-0.361,0.678) (0.165,-0.384) (-0.192,-0.878) (0.648,-0.035) (0.293,0.699) (0.222,-0.046) (0.541,0.729) (-0.143,0.120) (-0.068,0.340) (0.535,-0.604) (0.851,0.143) (0.154,0.172) (0.102,0.410) (-0.631,-0.087) (0.428,-0.306) (0.244,0.191) (-0.444,0.022) (-0.027,0.144) (0.395,0.743) (-0.183,-0.134) (-0.094,0.452) (0.507,-0.245) (-0.685,0.154) (0.228,0.294) (-0.323,0.545) (0.012,0.163) (-0.251,0.136) (-0.648,-0.057) (0.210,-0.597)};  \end{tikzpicture} }}\\
\midrule
\multirow{4}{*}{III}
 & $\#$PCs & 17 & 7 & 8 & 13 & 38 & 1 & 31\\
 & MNV & \uuline{<1e-08} & \uuline{<1e-08} & \uuline{<1e-08} & \uuline{<1e-08} & \uuline{<1e-08} & * & \uuline{<1e-08}\\
 & $t$-test & \uuline{<1e-08} & \uuline{<1e-08} & \uuline{<1e-08} & \uuline{<1e-08} & \uuline{<1e-08} & \uuline{<1e-08} & \uuline{<1e-08}\\
 & PCS & \raisebox{-.5\height}{\resizebox {1.2cm} {1.2cm} { \begin{tikzpicture}[scale=6] \path (-1.2,-1.2) (1.2,1.2);\draw[very thin,color=gray] (0,1.1)--(0,-1.1); \draw[very thin,color=gray] (1.1,0)--(-1.1,0); \path plot[mark=square*,mark options={color=red},mark size=0.8pt] coordinates { (-0.419,0.468) (-0.489,0.668) (-0.690,-0.147) (-0.630,0.139) (-0.704,-0.092) (-0.701,-0.180) (-0.568,0.334) (-1.000,-0.646) (-0.708,-0.084) (-0.462,0.058) (-0.821,-0.268) (-0.655,0.052) (-0.656,0.253) (-0.646,0.355) (-0.652,-0.075) (-0.570,0.371) (-0.815,-0.092) (-0.993,-0.257) (-0.606,0.258) (-0.685,-0.121) (-0.839,-0.310) (-0.770,-0.716) (-0.490,0.186) (-0.843,-0.211) (-0.377,0.880) (-0.733,0.048) (-0.609,0.234) (-0.558,0.397) (-0.794,0.078) (-0.699,0.202)};  \path plot[mark=diamond*,mark size=1pt] coordinates { (0.615,-0.498) (0.812,-0.124) (0.690,0.050) (0.778,0.265) (0.339,-0.590) (0.562,-0.365) (0.706,0.222) (0.535,-0.504) (0.668,0.229) (0.458,-0.111) (0.765,-0.143) (0.699,-0.230) (0.741,0.393) (0.872,0.346) (0.648,0.307) (0.828,0.451) (0.598,-0.395) (0.726,0.331) (0.828,0.285) (0.428,-0.506) (0.263,-0.722) (0.862,-0.152) (0.418,-0.191) (0.928,0.061) (0.770,0.206) (0.564,-0.497) (0.889,0.180) (0.777,0.125) (0.723,-0.165) (0.689,-0.038)};  \end{tikzpicture} }} & \raisebox{-.5\height}{\resizebox {1.2cm} {1.2cm} { \begin{tikzpicture}[scale=6] \path (-1.2,-1.2) (1.2,1.2);\draw[very thin,color=gray] (0,1.1)--(0,-1.1); \draw[very thin,color=gray] (1.1,0)--(-1.1,0); \path plot[mark=square*,mark options={color=red},mark size=0.8pt] coordinates { (-0.751,0.071) (-0.751,-0.042) (-0.903,0.066) (-0.873,0.013) (-0.917,-0.049) (-0.883,0.094) (-0.816,-0.033) (-1.000,0.007) (-0.878,-0.040) (-0.775,0.073) (-0.924,-0.039) (-0.876,0.101) (-0.831,-0.042) (-0.862,-0.142) (-0.865,0.084) (-0.797,-0.072) (-0.922,0.048) (-0.982,-0.007) (-0.842,0.021) (-0.855,-0.016) (-0.916,-0.040) (-0.990,-0.025) (-0.828,-0.027) (-0.958,-0.008) (-0.748,0.012) (-0.853,0.054) (-0.772,-0.102) (-0.842,-0.124) (-0.892,-0.029) (-0.870,0.220)};  \path plot[mark=diamond*,mark size=1pt] coordinates { (0.893,0.001) (0.870,-0.008) (0.867,0.027) (0.925,0.069) (0.762,0.139) (0.851,-0.379) (0.926,-0.171) (0.788,0.036) (0.892,-0.301) (0.804,-0.015) (0.867,0.111) (0.828,-0.272) (0.893,-0.374) (0.940,0.014) (0.940,0.247) (0.886,-0.400) (0.837,0.371) (0.927,0.154) (0.949,-0.002) (0.827,0.383) (0.676,-0.053) (0.930,0.270) (0.782,-0.244) (0.933,0.218) (0.907,0.172) (0.781,-0.012) (0.894,0.243) (0.879,-0.024) (0.829,-0.170) (0.889,-0.052)};  \end{tikzpicture} }} & \raisebox{-.5\height}{\resizebox {1.2cm} {1.2cm} { \begin{tikzpicture}[scale=6] \path (-1.2,-1.2) (1.2,1.2);\draw[very thin,color=gray] (0,1.1)--(0,-1.1); \draw[very thin,color=gray] (1.1,0)--(-1.1,0); \path plot[mark=square*,mark options={color=red},mark size=0.8pt] coordinates { (-0.711,0.104) (-0.652,0.228) (-0.850,-0.120) (-0.788,-0.059) (-0.860,0.017) (-0.806,-0.193) (-0.767,0.088) (-0.987,-0.175) (-0.805,-0.010) (-0.699,-0.104) (-0.864,-0.092) (-0.797,-0.113) (-0.767,0.037) (-0.795,0.390) (-0.798,-0.176) (-0.761,0.128) (-0.912,-0.021) (-0.963,0.042) (-0.762,0.101) (-0.816,-0.143) (-0.881,-0.098) (-1.000,-0.338) (-0.787,0.003) (-0.901,-0.064) (-0.634,0.342) (-0.807,-0.001) (-0.691,-0.099) (-0.747,0.267) (-0.877,0.091) (-0.805,0.150)};  \path plot[mark=diamond*,mark size=1pt] coordinates { (0.811,-0.267) (0.797,-0.108) (0.806,0.045) (0.884,0.059) (0.639,-0.103) (0.772,0.005) (0.847,0.175) (0.737,-0.077) (0.827,0.318) (0.761,0.203) (0.844,-0.153) (0.764,0.012) (0.888,0.280) (0.889,0.100) (0.876,-0.029) (0.859,0.502) (0.777,-0.336) (0.883,0.020) (0.941,-0.125) (0.699,-0.173) (0.569,-0.108) (0.876,-0.402) (0.735,0.041) (0.909,-0.183) (0.843,0.044) (0.700,-0.074) (0.860,0.088) (0.861,0.020) (0.759,0.153) (0.879,-0.110)};  \end{tikzpicture} }} & \raisebox{-.5\height}{\resizebox {1.2cm} {1.2cm} { \begin{tikzpicture}[scale=6] \path (-1.2,-1.2) (1.2,1.2);\draw[very thin,color=gray] (0,1.1)--(0,-1.1); \draw[very thin,color=gray] (1.1,0)--(-1.1,0); \path plot[mark=square*,mark options={color=red},mark size=0.8pt] coordinates { (-0.741,-0.270) (-0.688,-0.216) (-0.870,0.079) (-0.853,-0.002) (-0.924,0.048) (-0.886,0.084) (-0.768,-0.156) (-0.994,0.218) (-0.830,0.134) (-0.792,0.008) (-0.907,0.112) (-0.819,0.026) (-0.753,-0.132) (-0.837,-0.234) (-0.856,0.035) (-0.841,-0.111) (-0.882,0.099) (-0.937,0.076) (-0.757,-0.065) (-0.874,0.084) (-0.889,0.242) (-1.000,0.323) (-0.863,-0.118) (-0.931,0.146) (-0.733,-0.408) (-0.811,0.025) (-0.667,-0.015) (-0.810,-0.123) (-0.911,-0.055) (-0.857,-0.115)};  \path plot[mark=diamond*,mark size=1pt] coordinates { (0.873,0.308) (0.785,-0.013) (0.858,-0.051) (0.905,-0.104) (0.688,0.181) (0.870,0.317) (0.878,-0.080) (0.838,0.128) (0.831,-0.114) (0.900,0.072) (0.886,0.150) (0.781,0.114) (0.991,-0.205) (0.958,-0.204) (0.927,-0.170) (0.838,-0.245) (0.768,0.165) (0.837,-0.165) (0.945,-0.051) (0.794,0.153) (0.628,0.318) (0.856,0.120) (0.815,0.133) (0.849,-0.031) (0.867,-0.164) (0.763,0.130) (0.905,-0.312) (0.813,-0.128) (0.775,0.007) (0.859,0.026)};  \end{tikzpicture} }} & \raisebox{-.5\height}{\resizebox {1.2cm} {1.2cm} { \begin{tikzpicture}[scale=6] \path (-1.2,-1.2) (1.2,1.2);\draw[very thin,color=gray] (0,1.1)--(0,-1.1); \draw[very thin,color=gray] (1.1,0)--(-1.1,0); \path plot[mark=square*,mark options={color=red},mark size=0.8pt] coordinates { (-0.544,-0.107) (-0.870,-0.080) (-0.871,-0.213) (-0.694,0.018) (-0.605,0.055) (-0.597,-0.232) (-0.610,0.175) (-0.617,0.007) (-0.515,-0.152) (-0.332,-0.147) (-0.852,-0.082) (-0.708,0.030) (-0.743,0.097) (-0.773,0.571) (-0.665,-0.178) (-0.871,0.069) (-0.818,-0.084) (-0.830,-0.098) (-0.933,0.218) (-0.674,0.156) (-0.615,0.139) (-0.385,-0.345) (-0.568,-0.107) (-0.648,0.032) (-0.779,0.044) (-0.878,-0.116) (-0.820,0.056) (-0.958,0.405) (-0.835,0.156) (-0.605,-0.262)};  \path plot[mark=diamond*,mark size=1pt] coordinates { (0.730,-0.193) (0.622,-0.026) (0.619,0.048) (0.673,0.016) (0.743,-0.105) (0.884,0.509) (0.582,0.219) (0.510,-0.359) (0.824,0.419) (0.648,0.345) (0.637,-0.068) (0.781,0.145) (0.663,0.677) (0.537,0.014) (0.634,-0.297) (1.000,0.573) (0.583,-0.834) (0.674,-0.157) (0.717,0.005) (0.734,-0.423) (0.937,0.199) (0.825,-0.478) (0.695,0.412) (0.830,-0.210) (0.797,0.031) (0.657,-0.121) (0.816,-0.183) (0.622,-0.078) (0.730,0.079) (0.508,-0.186)};  \end{tikzpicture} }} & \raisebox{-.5\height}{\resizebox {1.2cm} {1.2cm} { \begin{tikzpicture}[scale=6] \path (-1.2,-1.2) (1.2,1.2);\draw[very thin,color=gray] (0,1.1)--(0,-1.1); \draw[very thin,color=gray] (1.1,0)--(-1.1,0); \path plot[mark=square*,mark options={color=red},mark size=0.8pt] coordinates { (-0.874,0.094) (-0.841,0.132) (-0.929,-0.046) (-0.901,-0.022) (-0.933,-0.009) (-0.904,-0.087) (-0.897,0.052) (-0.976,-0.133) (-0.904,-0.028) (-0.868,-0.039) (-0.928,-0.077) (-0.903,-0.034) (-0.889,0.015) (-0.902,0.167) (-0.904,-0.070) (-0.893,0.070) (-0.951,-0.017) (-0.964,-0.014) (-0.889,0.064) (-0.912,-0.080) (-0.935,-0.087) (-1.000,-0.202) (-0.911,0.014) (-0.943,-0.053) (-0.840,0.225) (-0.903,-0.000) (-0.854,-0.053) (-0.885,0.122) (-0.935,0.034) (-0.905,0.112)};  \path plot[mark=diamond*,mark size=1pt] coordinates { (0.912,-0.148) (0.894,-0.044) (0.906,0.032) (0.935,0.061) (0.850,-0.056) (0.897,-0.078) (0.922,0.063) (0.884,-0.056) (0.918,0.106) (0.901,0.070) (0.918,-0.047) (0.887,-0.042) (0.942,0.093) (0.934,0.072) (0.943,0.038) (0.923,0.186) (0.898,-0.104) (0.939,0.052) (0.959,-0.024) (0.875,-0.051) (0.824,-0.091) (0.927,-0.127) (0.890,-0.032) (0.938,-0.029) (0.919,0.061) (0.867,-0.062) (0.920,0.097) (0.928,0.016) (0.884,0.037) (0.940,-0.045)};  \end{tikzpicture} }} & \raisebox{-.5\height}{\resizebox {1.2cm} {1.2cm} { \begin{tikzpicture}[scale=6] \path (-1.2,-1.2) (1.2,1.2);\draw[very thin,color=gray] (0,1.1)--(0,-1.1); \draw[very thin,color=gray] (1.1,0)--(-1.1,0); \path plot[mark=square*,mark options={color=red},mark size=0.8pt] coordinates { (-0.237,0.056) (-0.167,0.294) (-0.597,-0.219) (-0.375,-0.103) (-0.572,0.060) (-0.583,-0.282) (-0.316,0.123) (-1.000,-0.138) (-0.533,0.021) (-0.222,-0.200) (-0.734,-0.038) (-0.497,-0.246) (-0.408,0.039) (-0.480,0.690) (-0.522,-0.327) (-0.366,0.180) (-0.699,-0.038) (-0.874,0.098) (-0.419,0.143) (-0.550,-0.167) (-0.740,-0.011) (-0.763,-0.410) (-0.263,-0.116) (-0.748,-0.001) (-0.022,0.361) (-0.551,-0.015) (-0.333,-0.146) (-0.365,0.433) (-0.645,0.172) (-0.554,0.085)};  \path plot[mark=diamond*,mark size=1pt] coordinates { (0.432,-0.313) (0.612,-0.211) (0.500,0.057) (0.743,-0.006) (0.108,-0.078) (0.419,0.226) (0.666,0.281) (0.195,-0.058) (0.534,0.560) (0.221,0.362) (0.633,-0.274) (0.472,0.118) (0.648,0.523) (0.751,0.068) (0.579,-0.177) (0.684,0.851) (0.400,-0.647) (0.701,-0.120) (0.792,-0.303) (0.161,-0.277) (-0.064,0.015) (0.789,-0.737) (0.253,0.194) (0.806,-0.405) (0.611,0.002) (0.235,-0.020) (0.666,-0.038) (0.628,-0.000) (0.429,0.300) (0.528,-0.191)};  \end{tikzpicture} }}\\
\bottomrule
\end{tabular}
} 
\caption{Model-independent comparison for model size 400, parameter set 1, and $n=30$ runs per configuration. Case I weights two similar setups, case II compares setups with a small implementation difference, and case III compares setups with a different parameter. $\#$PCs is the number of PCs which explain at least 90\% of variance, while MNV and $t$-test refer to the $p$-value yielded by the MANOVA ($\#$PCs) and $t$ (first PC) hypothesis tests, respectively. PCS shows the representation of the respective output in the first two dimensions of the PC space. Output $\widetilde{A}$ refers to the concatenation of all outputs (range scaled). The MANOVA test is not applicable to one PC, and the respective $p$-values are replaced with an asterisk (*). $P$-values lower than $0.05$ are underlined, while $p$-values lower than $0.01$ are double-underlined.}
\label{tab:micomps}
\end{center}
\end{table}

\subsection{Model-independent selection of FMs}
\label{sec:results:micomp}

Table~\ref{tab:micomps} presents the results for the model-independent comparison approach for $400@1$. Regarding the number of PCs, there is not much difference between case I and II, except for the mean prey energy, $\mean{E}^s$, as this output is clearly the most affected by the introduced change (no agent list shuffling). For case III, however, less PCs are required to explain the same percentage of variance in all outputs, meaning that configurations 1 and 4 are less aligned than configurations 1 and 2 (case I) and configurations 1 and 3 (case II). Nonetheless, the $p$-values offer more objective answers, displaying increasingly significant differences from case I to case III, i.e., in line with the results from the empirical approach. $P$-values were obtained with the $t$-test for the first PC, and with MANOVA for the number of PCs which explain at least 90\% of the variance. The assumptions for these parametric tests seem to be verified, and are discussed with additional detail in Section~\ref{sec:results:assumptions}. MANOVA appears to be more sensitive to implementation differences than the $t$-test, generally presenting more significant $p$-values. As would be expected, significant $p$-values from both MANOVA and $t$-tests are associated with clearer distinctions between samples in the PC scatter plots.

The concatenated output, $\widetilde{A}$, has less discriminatory power than individual outputs. This is especially clear for case II, where both the $t$-test and scatter plot do not suggest significant discrepancies. However, the MANOVA test is able to pick up the implementation difference, yielding a significant $p$-value. While in this case MANOVA on $\widetilde{A}$ answers to the question of whether the implementations are statistically equivalent, only a comprehensive analysis of individual outputs allows to diagnose how the model is affected by the implementation differences. For example, the number of PCs and the scatter plot indicate that the $\mean{E}^s$ output is by far the most affected in case II, something which is not possible to deduce by just analyzing $\widetilde{A}$. This consideration would also be difficult to infer from the empirical comparison data in Table~\ref{tab:mdcomps}, as the $p$-values are not much different from those associated with other outputs. As described in Section \ref{sec:micomp:multout}, outputs were range scaled prior to concatenation. We have experimented with other types of centering and scaling \cite{berg2006centering}, and did not find major differences on how the proposed model-independent comparison method evaluates the concatenated output.

Results for the remaining size/set combinations, provided in Supplementary Tables S2.1\textendash{}S2.8, are in accordance with the discussion thus far. The number of PCs required to explain 90\% of the variance for cases II and III are generally smaller than for case I. Furthermore, the number of PCs for parameter set 2 is consistently lower in case II than in case III, attesting to what was observed in the model-dependent analysis: configurations compared in case II are more dissimilar than those in case III. This is also corroborated by the $p$-values, and even more clearly, by the PC scatter plots. For the majority of individual comparisons in cases II and III, MANOVA seems more sensitive to implementation differences than the $t$-test. However, there are some instances where the latter is able to perform better distinctions, as for example in case II of the $100@1$ size/set combination. Concerning the concatenated output, it was generally adequate for detecting implementation differences, with MANOVA and/or the $t$-test yielding significant $p$-values. Nonetheless, this approach failed for case III of $100@2$. Analyzing the overall results for all size/set combinations, it is also possible to conclude that, as model size increases, implementation differences become more pronounced for both parameter sets.

\begin{table}[!htbp]
\begin{center}
\resizebox{\columnwidth}{!}{%
\begin{tabular}{clrrrrrrr}
\toprule
\multirow{2}{*}{Comp.} & \multirow{2}{*}{Data} & \multicolumn{7}{c}{Outputs} \\
\cmidrule(l){3-9}
 &  & $P^s$ & $P^w$ & $P^c$ & $\mean{E}^s$ & $\mean{E}^w$ & $\mean{C}$ & $\widetilde{A}$\\
\midrule
\multirow{12}{*}{I}
 & \% var. (PC1) &   19.1 &   13.0 &   15.9 &   11.7 &    8.2 &   15.8 &   10.9\\
 & \% var. (PC2) &    8.6 &    7.5 &    8.7 &    6.4 &    6.2 &    8.6 &    6.4\\
 & \% var. (PC3) &    7.1 &    6.8 &    7.5 &    5.6 &    4.5 &    7.4 &    6.1\\
 & \% var. (PC4) &    6.4 &    6.5 &    6.6 &    5.3 &    4.3 &    6.6 &    5.8\\
\cmidrule(l){2-9}
 & $t$-test (PC1) & 0.530 & 0.836 & 0.804 & 0.784 & 0.378 & 0.805 & 0.879\\
 & $t$-test (PC2) & \uline{0.021} & 0.057 & \uline{0.012} & 0.115 & 0.915 & \uline{0.012} & \uline{0.041}\\
 & $t$-test (PC3) & 0.319 & 0.608 & 0.491 & 0.081 & 0.372 & 0.496 & 0.389\\
 & $t$-test (PC4) & 0.763 & 0.200 & 0.872 & 0.473 & 0.327 & 0.868 & 0.524\\
\cmidrule(l){2-9}
 & $t$-test* (PC1) & 1.000 & 1.000 & 1.000 & 1.000 & 1.000 & 1.000 & 1.000\\
 & $t$-test* (PC2) & 0.240 & 0.757 & 0.135 & 1.000 & 1.000 & 0.136 & 0.633\\
 & $t$-test* (PC3) & 1.000 & 1.000 & 1.000 & 1.000 & 1.000 & 1.000 & 1.000\\
 & $t$-test* (PC4) & 1.000 & 1.000 & 1.000 & 1.000 & 1.000 & 1.000 & 1.000\\
\midrule
\multirow{12}{*}{II}
 & \% var. (PC1) &   20.1 &   19.3 &   16.4 &   77.2 &    8.0 &   16.2 &   11.4\\
 & \% var. (PC2) &   10.7 &   10.5 &   11.1 &    2.6 &    6.1 &   11.0 &    8.8\\
 & \% var. (PC3) &    7.0 &    8.0 &    7.3 &    1.9 &    4.7 &    7.2 &    6.2\\
 & \% var. (PC4) &    5.9 &    5.3 &    6.3 &    1.3 &    4.1 &    6.3 &    5.5\\
\cmidrule(l){2-9}
 & $t$-test (PC1) & \uline{0.042} & \uuline{<1e-08} & 0.108 & \uuline{<1e-08} & \uline{0.017} & 0.109 & 0.390\\
 & $t$-test (PC2) & 0.128 & \uline{0.029} & 0.106 & 0.720 & 0.175 & 0.107 & 0.086\\
 & $t$-test (PC3) & \uuline{0.004} & 0.110 & \uuline{0.006} & 0.885 & 0.909 & \uuline{0.006} & \uuline{0.002}\\
 & $t$-test (PC4) & 0.767 & 0.401 & 0.966 & 0.978 & 0.060 & 0.954 & 0.631\\
\cmidrule(l){2-9}
 & $t$-test* (PC1) & 0.209 & \uuline{<1e-08} & 0.661 & \uuline{<1e-08} & 0.211 & 0.669 & 1.000\\
 & $t$-test* (PC2) & 1.000 & 0.276 & 0.955 & 1.000 & 1.000 & 0.971 & 0.975\\
 & $t$-test* (PC3) & 0.055 & 1.000 & 0.079 & 1.000 & 1.000 & 0.077 & \uline{0.028}\\
 & $t$-test* (PC4) & 1.000 & 1.000 & 1.000 & 1.000 & 1.000 & 1.000 & 1.000\\
\midrule
\multirow{12}{*}{III}
 & \% var. (PC1) &   38.6 &   79.2 &   74.3 &   73.6 &   32.4 &   94.0 &   29.1\\
 & \% var. (PC2) &    9.4 &    2.6 &    3.4 &    2.8 &    4.4 &    0.8 &    7.9\\
 & \% var. (PC3) &    7.6 &    2.4 &    3.2 &    2.4 &    4.2 &    0.7 &    7.7\\
 & \% var. (PC4) &    6.9 &    2.0 &    2.7 &    2.0 &    3.7 &    0.6 &    5.7\\
\cmidrule(l){2-9}
 & $t$-test (PC1) & \uuline{<1e-08} & \uuline{<1e-08} & \uuline{<1e-08} & \uuline{<1e-08} & \uuline{<1e-08} & \uuline{<1e-08} & \uuline{<1e-08}\\
 & $t$-test (PC2) & 0.183 & 0.966 & 0.792 & 0.662 & 0.979 & 0.878 & 0.792\\
 & $t$-test (PC3) & 0.734 & 0.906 & 0.826 & 0.976 & 0.777 & 0.951 & 0.754\\
 & $t$-test (PC4) & 0.602 & 0.800 & 0.587 & 0.827 & 0.979 & 0.875 & 0.555\\
\cmidrule(l){2-9}
 & $t$-test* (PC1) & \uuline{<1e-08} & \uuline{<1e-08} & \uuline{<1e-08} & \uuline{<1e-08} & \uuline{<1e-08} & \uuline{<1e-08} & \uuline{<1e-08}\\
 & $t$-test* (PC2) & 1.000 & 1.000 & 1.000 & 1.000 & 1.000 & 1.000 & 1.000\\
 & $t$-test* (PC3) & 1.000 & 1.000 & 1.000 & 1.000 & 1.000 & 1.000 & 1.000\\
 & $t$-test* (PC4) & 1.000 & 1.000 & 1.000 & 1.000 & 1.000 & 1.000 & 1.000\\
\bottomrule
\end{tabular}
} 
\caption{Percentage of explained variance (\% var.) and $t$-test $p$-values before ($t$-test) and after ($t$-test*) weighted Bonferroni correction of the first four PCs for model size 400, parameter set 1, and $n=30$ runs per configuration. Case I weights two similar setups, case II compares setups with a small implementation difference, and case III compares setups with a different parameter. $P$-values lower than $0.05$ are underlined, while $p$-values lower than $0.01$ are double-underlined.}
\label{tab:micomps_tpcs}
\end{center}
\end{table}

\subsubsection{Assessing the $t$-test applied to individual PCs and the distribution of explained variance}

Table~\ref{tab:micomps_tpcs} shows, for the first four PCs of the $400@1$ combination, the percentage of variance explained by individual PCs, as well as the respective $t$-test $p$-values, before and after weighted Bonferroni correction. For case I, $p$-values for the first PC are never significant, but a few for the second PC are significant at the $\alpha=0.05$ level. However, no $p$-values remain so after adjustment with the weighted Bonferroni correction. For case II, $p$-values for the first PC are usually the most significant, although this does not always hold, e.g., output $\widetilde{A}$. In this instance, only the 3\textsuperscript{rd}, 6\textsuperscript{th} and 15\textsuperscript{th} PC $p$-values are significant (the 3\textsuperscript{rd} and 6\textsuperscript{th} remain so at $\alpha=0.05$ after correction), which is the reason for MANOVA catching the implementation difference (as shown in Table~\ref{tab:micomps}). For case III, the first PC $p$-value is always highly significant (before and after weighted Bonferroni correction), with the PC2\textendash{}PC4 $p$-values showing no significance at all. The implementation differences in the first PC, captured by the $t$-test, seem sufficient for MANOVA, which considers all PCs simultaneously, to also spot the dissimilarities. The distribution of explained variance along PCs also reflects what was discussed in Section \ref{sec:micomp}. More specifically, it is possible to observe that the first PC or PCs explain more variance in misaligned cases II and III than in case I (aligned).

These observations generally remain valid when considering all tested size/set combinations (Tables S3.1\textendash{}S3.8, provided as Supplementary material). For case I, in which the compared configurations are assumed to be aligned, a few significant PC1 $p$-values (at the $\alpha=0.05$ level) stand-out for $100@1$. Nonetheless, after the weighted Bonferroni correction, all $t$-test $p$-values from PC1\textendash{}PC4 are non-significant for all size/set combinations. For case II, a global size/set analysis confirms that $p$-values for the first PC are usually the most significant for all outputs. The most notable exception occurs for the $200@1$ combination, in which only the $\mean{E}^s$ output presents a significant $p$-value (at $\alpha=0.01$, before and after weighted Bonferroni correction). However, this $p$-value is enough for the respective MANOVA (Table~S2.3) to catch the difference directly in $\mean{E}^s$, and indirectly in the concatenated output. With very few exceptions, the PC2\textendash{}PC4 $p$-values are not significant, especially after the weighted Bonferroni correction. Results for case III broadly confirm what was observed for $400@1$, i.e., that the first PC $p$-value is always highly significant (before and after weighted Bonferroni correction), with the PC2\textendash{}PC4 $p$-values showing little to no significance. The exception is $100@2$, for which only the PC1 $p$-value for $P^s$ is significant at $\alpha=0.05$, but losing significance after the weighted Bonferroni correction. Nonetheless, dissimilarities between configurations 1 and 4 are detected by the MANOVA test (Table~S2.2) in the majority of outputs due to misalignments in PCs other than the first (e.g., for output $\mean{C}$ these are visible for PC2 and PC3). Considering these results, it is possible to conclude that the $p$-value of the first PC is undoubtedly the most important for evaluating model alignment. Nonetheless, there are also instances, namely for smaller model sizes, where it is necessary to look at the MANOVA $p$-value in order to draw solid conclusions. The distribution of explained variance along PCs generally reflects the alignment of configurations, with less aligned ones having more variance explained by the first PC or PCs. Again the exception is $100@2$, where the difference in explained variance does not change much from case to case.

\begin{table}[!htbp]
\begin{center}
\resizebox{\columnwidth}{!}{%
\begin{tabular}{clrrrrrrr}
\toprule
\multirow{2}{*}{Comp.} & \multirow{2}{*}{Data} & \multicolumn{7}{c}{Outputs} \\
\cmidrule(l){3-9}
 &  & $P^s$ & $P^w$ & $P^c$ & $\mean{E}^s$ & $\mean{E}^w$ & $\mean{C}$ & $\widetilde{A}$\\
\midrule
\multirow{8}{*}{I}
 & $\#$PCs (30\% var.) & 3 & 4 & 3 & 5 & 6 & 3 & 5\\
 & $\#$PCs (50\% var.) & 6 & 8 & 6 & 10 & 13 & 7 & 9\\
 & $\#$PCs (70\% var.) & 11 & 15 & 12 & 19 & 25 & 12 & 19\\
 & $\#$PCs (90\% var.) & 24 & 32 & 25 & 36 & 43 & 26 & 39\\
\cmidrule(l){2-9}
 & MNV (30\% var.) & 0.080 & 0.235 & 0.078 & 0.185 & 0.696 & 0.078 & 0.297\\
 & MNV (50\% var.) & 0.267 & 0.514 & 0.278 & \uline{0.020} & 0.530 & 0.311 & 0.614\\
 & MNV (70\% var.) & 0.241 & 0.611 & 0.344 & \uline{0.044} & 0.682 & 0.344 & 0.527\\
 & MNV (90\% var.) & 0.528 & 0.258 & 0.548 & 0.105 & 0.746 & 0.577 & 0.704\\
\midrule
\multirow{8}{*}{II}
 & $\#$PCs (30\% var.) & 2 & 3 & 3 & 1 & 6 & 3 & 4\\
 & $\#$PCs (50\% var.) & 6 & 6 & 6 & 1 & 13 & 6 & 9\\
 & $\#$PCs (70\% var.) & 11 & 12 & 11 & 1 & 24 & 12 & 18\\
 & $\#$PCs (90\% var.) & 24 & 29 & 24 & 12 & 43 & 25 & 38\\
\cmidrule(l){2-9}
 & MNV (30\% var.) & \uline{0.038} & \uuline{<1e-08} & \uuline{0.004} & * & \uline{0.035} & \uuline{0.004} & \uuline{0.007}\\
 & MNV (50\% var.) & \uuline{2e-06} & \uuline{<1e-08} & \uuline{3e-04} & * & \uline{0.034} & \uuline{3e-04} & \uuline{3e-04}\\
 & MNV (70\% var.) & \uuline{4e-05} & \uuline{<1e-08} & \uuline{0.002} & * & \uline{0.049} & \uuline{0.001} & \uuline{2e-06}\\
 & MNV (90\% var.) & \uuline{<1e-08} & \uuline{<1e-08} & \uuline{5e-08} & \uuline{<1e-08} & 0.467 & \uuline{1e-07} & \uuline{<1e-08}\\
\midrule
\multirow{8}{*}{III}
 & $\#$PCs (30\% var.) & 1 & 1 & 1 & 1 & 1 & 1 & 2\\
 & $\#$PCs (50\% var.) & 3 & 1 & 1 & 1 & 6 & 1 & 4\\
 & $\#$PCs (70\% var.) & 6 & 1 & 1 & 1 & 17 & 1 & 11\\
 & $\#$PCs (90\% var.) & 17 & 7 & 8 & 13 & 38 & 1 & 31\\
\cmidrule(l){2-9}
 & MNV (30\% var.) & * & * & * & * & * & * & \uuline{<1e-08}\\
 & MNV (50\% var.) & \uuline{<1e-08} & * & * & * & \uuline{<1e-08} & * & \uuline{<1e-08}\\
 & MNV (70\% var.) & \uuline{<1e-08} & * & * & * & \uuline{<1e-08} & * & \uuline{<1e-08}\\
 & MNV (90\% var.) & \uuline{<1e-08} & \uuline{<1e-08} & \uuline{<1e-08} & \uuline{<1e-08} & \uuline{<1e-08} & * & \uuline{<1e-08}\\
\bottomrule
\end{tabular}
} 
\caption{Number of principal components ($\#$PCs) and MANOVA test $p$-values (MNV) for several percentages of explained variance for model size 400, parameter set 1, and $n=30$ runs per configuration. Case I weights two similar setups, case II compares setups with a small implementation difference, and case III compares setups with a different parameter. Output $\widetilde{A}$ refers to the concatenation of all outputs (range scaled). The MANOVA test is not applicable to one PC, and the respective $p$-values are replaced with an asterisk (*). $P$-values lower than $0.05$ are underlined, while $p$-values lower than $0.01$ are double-underlined.}
\label{tab:micomps_multvarexp}
\end{center}
\end{table}

\subsubsection{Changing the variance for the selection of the number of PCs for MANOVA}

The MANOVA $p$-values shown in Table~\ref{tab:micomps} were obtained by requiring that the respective number of PCs explain at least 90\% of the variance in the data. As this percentage must be predefined by the model developer, it is important to understand how the MANOVA results are affected when this value is changed. Table~\ref{tab:micomps_multvarexp} shows, for the $400@1$ instance, the number of PCs, as well as the associated MANOVA $p$-values, required to explain 30\%, 50\%, 70\% and 90\% of the variance. Three aspects can be highlighted from this table: 1) conclusions concerning configuration alignment do not change with different percentages of variance; 2) more variance seems to make the MANOVA test more sensitive in case II, although this trend is not very well-defined; 3) lower prespecified percentages of variance imply less PCs to explain it; in the limit, this makes the MANOVA test inapplicable if just one PC is required, as observed for several instances of cases II and III. This latter aspect is not crucial, because the $t$-test is the equivalent of a ``univariate MANOVA'' for two groups. Nonetheless, if the MANOVA $p$-values are deemed important for a specific comparison, it seems preferable to specify a higher percentage of variance to explain.

Results for the remaining size/set combinations, provided in Supplementary Tables~S4.1\textendash{}S4.8, generally follow the tendencies verified for $400@1$, perhaps further blurring the slight trend observed in the relation between percentage of variance and MANOVA sensitivity.

\subsubsection{Assumptions for the $t$ and MANOVA parametric tests}
\label{sec:results:assumptions}

As described in Section \ref{sec:micomp}, the $t$ and MANOVA tests make several assumptions about the underlying data. To confirm that the presented results are statistically valid, it is important to perform an overall survey of these assumptions, namely: 1) whether the data is normally distributed within groups; and, 2) if samples are drawn from populations with equal variances. The former assumption can be verified with the Shapiro\textendash{}Wilk test for univariate normality or the Royston test for multivariate normality. The latter can be evaluated with the Bartlett test or Box's M test for the univariate and multivariate cases, respectively. Table \ref{tab:micomps_assumpt} presents aggregated results for these tests for all size/set combinations. More specifically, this table shows the percentage of $p$-values from the specified tests which fall within the non-significant ($p>0.05$) and significant ($0.01<p<0.05$ and $p<0.01$) intervals.

\begin{table}[ht]
\begin{center}
\begin{tabular}{lrrr}
\toprule
Test & \multicolumn{1}{c}{$p>0.05$} & \multicolumn{1}{c}{$0.01<p<0.05$} & \multicolumn{1}{c}{$p<0.01$} \\
\midrule
SW (PC1)       & 94.77\% (95.24\%)  &  4.19\% (4.46\%) &  1.04\% (0.30\%) \\
Bartlett (PC1) & 92.65\% (89.29\%)  &  5.15\% (8.93\%) &  2.20\% (1.79\%) \\
Royston       & 88.51\%  &  6.52\% &  4.97\% \\
Box's M       & 27.33\%  & 16.15\% & 56.52\% \\
\bottomrule
\end{tabular}
\caption{\label{tab:micomps_assumpt}Percentage of $p$-values yielded by the specified tests of assumptions which fall within the non-significant ($p>0.05$) and significant ($0.01<p<0.05$ and $p<0.01$) intervals. For the Shapiro\textendash{}Wilk (SW) and Bartlett tests, percentages outside of parenthesis refer to tests of individual groups for each PC, while percentages within parentheses refer to tests of individual groups for the first PC. These percentages consider all tested size/set combinations, for $n=30$ runs per configuration, and variance to explain prespecified at 90\%.}
\end{center}
\end{table}

The Shapiro\textendash{}Wilk test does not reject univariate normality in approximately 95\% of samples used for the $t$-tests. Of the remaining 5\%, only 1\% show significant rejection at the $\alpha=0.01$ level. This percentage is even lower for PC1 samples, the most critical for evaluating configuration alignment using the univariate approach. Samples also seem to generally have similar variance, as denoted by the 90\% of non-significant results for the Bartlett test. Multivariate normality is also verified by the Royston test in most instances. Since $p$-values are the probability of obtaining a result at least as extreme than what was actually observed (assuming that the null hypothesis is true), type I error rates are close to what would be expected, namely 95\% for $p>0.05$, 4\% for $0.01<p<0.05$ and 1\% for $p<0.01$. As such, the assumptions tested by the Shapiro\textendash{}Wilk, Bartlett and Royston tests seem to be generally verified. The same does not appear to hold for the MANOVA assumption of homogeneity of variance\textendash{}covariance matrices, evaluated with Box's M test. However, as stated in Section \ref{sec:tests}, this test is extremely sensitive and MANOVA is resilient to violations in this assumption when samples have equal sizes, which is the case here.

\subsection{Comparison performance with different sample sizes}

Larger sample sizes make hypothesis test more powerful, i.e., more likely to reject the null hypothesis when it is false, thus decreasing the probability of committing a type II error. 
Here we are essentially interested in determining, within the discussed frameworks, if and how conclusions concerning configuration alignment change with sample size. For this purpose, the empirical and the proposed model-independent comparison methods were tested with samples of $n=10$ and $n=100$ replicates for the $400@1$ combination. The comparison with samples of size 10 used the first 10 observations from the $n=30$ runs per configuration setup, while the comparison with samples of size 100 used a new set of $n=100$ runs per configuration. Results from this analysis are provided in Supplementary Tables S5.1 (empirical approach, 10 runs), S5.2 (model-independent approach, 10 runs), S6.1 (empirical approach, 100 runs) and S6.2 (model-independent approach, 100 runs).

Using samples of 10 observations, the empirical selection of FMs still manages to find configuration differences in cases II and III, although not as clearly as for $n=30$. In case II, the difference is found on the steady-state mean of the $P^w$ and $\mean{E}^s$ outputs. With $n=30$, differences are detected in several more FMs. The proposed model-independent comparison approach is also able to spot differences in cases II and III, likewise detecting differences in the $P^w$ and $\mean{E}^s$ outputs for case II. 

For $n=100$, both methods confirm what was observed for smaller sample sizes, although results are more pronounced, with very significant $p$-values for cases II and III. Two details emerge, however. First, for case II, the empirical approach hardly recognizes configuration differences for the mean predator energy output, $\mean{E}^w$. Such differences are exposed by the proposed model-independent method, both by the $t$-test and MANOVA. Second, the model-independent method perceives a difference where there should not be any, namely in output $\mean{E}^s$ of case I, via the MANOVA test. The PC1 $t$-test $p$-value is not significant, as is the case for the vast majority of remaining PCs. For the 72 PCs (dimensions) considered by MANOVA, the $t$-test $p$-value is significant at $\alpha=0.01$ in three cases: PC11, PC22 and PC48. After weighted Bonferroni correction, only the PC11 $p$-value, associated with 2.4\% of the variance, remains significant, with a value of $0.006$. Since these PCs, and PC11 in particular, account for a small amount of output variability, conclusions concerning output similarity (and consequently, model alignment), do not necessarily change. In practice this result is not totally unexpected, since the increase in power (due to a larger sample size) together with a MANOVA test on many PCs or dimensions, leads to a higher probability of detecting smaller or irrelevant differences. Overall, what seems clear for a sample of $n=100$ replicates, is that a $t$-test on the first PC of each individual output (or even the concatenated output) is enough to pick up implementation differences.

\section{Discussion}
\label{sec:discussion}

The empirical approach and the proposed model-independent method were able to identify the changes introduced in configurations 3 and 4, even though these were of different nature. The change introduced in configuration 3, case II, confirms the sensitivity of this type of models: a trivial implementation discrepancy led to statistically different behavior. In turn, the change in configuration 4, case III, was expected to be noticed, since it implied the alteration of a model parameter. Nonetheless, in specific conditions, namely for parameter set 2, configuration 3 presented more dissimilar behavior than configuration 4, when compared to configuration 1. Since parameter set 2 generates more agents during simulations (as described in Section \ref{sec:simmod:params}), the effect of not shuffling the agent list in configuration 3, case II, may be more pronounced. Additionally, the $c_r$ parameter is larger in parameter set 2 than in parameter set 1. As such, the smaller relative difference of decreasing $c_r$ by one unit in configuration 4 may lead to less significant discrepancies in case III.

In the context of a smaller sample size, namely for $n=10$ runs per configuration, both methods were capable of discerning differences in cases II and III. Differentiation was better with increasing number of replicates, i.e., when going from $n=10$, then $n=30$, and finally $n=100$. In practice, however, more replicates did not yield different conclusions regarding output and configuration alignment.

Globally, results for the proposed model-independent method were in accordance with the empirical approach on an per output basis. The method was also able to expose a few implementation differences not detected when empirically selecting FMs, and, in other instances, it identified these differences more conclusively. As such, the proposed method was shown to be a more direct, and often more effective alternative to the empirical and model-dependent selection of FMs.

The $p$-values yielded by the $t$ and MANOVA tests offer an objective assessment concerning the alignment of individual outputs. While in most instances the two tests generated the same decision, there were situations in which results differed. In case of alignment, the $t$-test on the first PC never produced $p$-values significant at $\alpha=0.01$, even prior to weighted Bonferroni correction. MANOVA, however, signaled a few false positives for the $\mean{E}^s$ output, especially in the $n=100$ observations per sample case, which also affected the $\widetilde{A}$ output. A careful analysis revealed the cause to be a misalignment in one or more PCs, which in themselves did not account for much output variance, but were sufficient to provoke a type I error in the MANOVA test. In case of configuration misalignment, MANOVA was generally more sensitive than the PC1 $t$-test, and with the exception of $100@1$, flagged a few misalignments unnoticed by the $t$-test. 

The concatenated output provides a single answer on whether two or more model implementations are aligned or not. In terms of false positives on $\widetilde{A}$, i.e., reporting a misalignment where none is expected, this was observed once by a MANOVA test for the $n=100$ runs per configuration instance. The $t$-test, however, did not flag any configuration difference. False negatives, i.e., failure to detect output mismatch when it is known to exist, occurred a few times for both MANOVA and the $t$-test, and were essentially associated with smaller model sizes. Additionally, neither test was able to detect differences on $\widetilde{A}$ in case II for the $n=10$ runs per configuration instance. 

\section{Recommendations on using the model-independent method}
\label{sec:recomend}

Given the results presented and discussed in the previous sections, a number of recommendations on using the proposed model-independent method can be made.

The $t$ and MANOVA tests should both be performed, and in case of disagreement, it is important to find the cause. More specifically, 1) if the PC1 $t$-test $p$-value is significant (and the MANOVA $p$-value is not), it should be adjusted with the weighted Bonferroni correction, and considered instead; 2) if the MANOVA $p$-value is significant (and the PC1 $t$-test $p$-value is not), the $t$-test should be performed on other PCs to find which ones are likely to be influencing the MANOVA result. A decision on alignment should consider how much variance the culprit PCs explain. As an alternative to performing the $t$-test on PC1 and MANOVA on the number of PCs which explain a minimum percentage of variance, it is also possible to perform the $t$-test on all PCs individually, and consider the resulting $p$-values after weighted Bonferroni correction. This also has the advantage of automatically privileging the most important PCs, i.e., the ones that explain more variance. Furthermore, if samples do not appear to be drawn from a normal distribution, a non-parametric test can be used instead of the $t$-test (e.g., Mann\textendash{}Whitney or Kolmogorov\textendash{}Smirnov).

The prespecified variance determines the number of PCs used in the MANOVA test. While conclusions concerning configuration alignment do not appear to change dramatically with different percentages of variance, it seems preferable to specify more variance than less, otherwise one may not be able to use MANOVA. A minimum variance of 90\% seems appropriate for this purpose, but this may vary with models,  outputs characteristics and parameterizations. Another possibility is to use a fixed number of PCs for MANOVA instead of specifying a percentage of variance to explain. A potential problem with this approach is that MANOVA could eventually consider very different percentages of variance from output to output and/or case to case. 

The eigenvalue structure, reflected in the number of PCs required to explain a prespecified percentage of variance, as well as in the variance explained by the first PC or PCs, is also an interesting comparison metric to consider. For aligned outputs, in contrast with misaligned ones, more PCs are required to explain the same amount of variance, and less variance is explained by the first PC or PCs. However, the information provided by this metric in general, and by the number of PCs in particular, is only useful when assessing multiple cases, for which it can be relatively compared. 

The scatter plot of the first two PC dimensions is also useful for quickly assessing possible output misalignments. If the outputs have a clear mismatch, points associated with runs from different implementations form visibly distinct groups. Although this technique is basically a type of face validation, it provides a useful intuition on the alignment of two or more implementations for any given output.

Finally, output concatenation also provides a practical and rapid way of assessing model alignment, but should not be fully relied upon to make a final decision. Additionally, the analysis of individual outputs allows to diagnose which model aspects are most affected by possible mismatches, something which the concatenated output does not permit.

Altogether, no single metric of the proposed method provides a definitive answer, and decisions concerning model alignment should be taken based on the different types of information it produces. 

\section{Conclusions}
\label{sec:conclusions}

A technique for comparing simulation models in a model-independent fashion was presented in this paper. The method dispenses the need for selecting model-specific FMs, and is appropriate for comparing multiple implementations of a conceptual model or replications of an existing model. The technique was validated against the commonly used approach of empirically selecting output statistical summaries using the PPHPC ABM on a number of model sizes and parameter sets. Results showed that the proposed method offered similar conclusions concerning output and model alignment, providing stronger misalignment detection power in some situations.

Although the proposed framework is presented in the context of simulation output comparison, it is a generic statistical comparison methodology. As such, it can be employed in circumstances where observations yield a large number of correlated variables \cite{fachada2016micompr}. 

\section*{Acknowledgments}

This work was supported by the Fundação para a Ciência e a Tecnologia (FCT) projects UID/\allowbreak EEA/\allowbreak 50009/\allowbreak 2013 and UID/\allowbreak MAT/\allowbreak 04561/\allowbreak 2013, and partially funded with grant SFRH/\allowbreak BD/\allowbreak 48310/\allowbreak 2008, also from FCT.

\bibliographystyle{elsarticle-num}

\includepdf[pages=1-37]{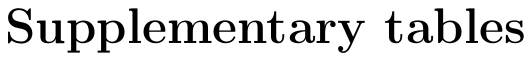}

\end{document}